\documentclass[11pt,a4paper]{article}
\usepackage{jheppub,bm,color}
\usepackage[dvipsnames]{xcolor}
\usepackage{hyperref}
\usepackage{amsmath,amssymb,bm}

\newcommand{\MIS}{MIS${}^*$~}
\newcommand{\tR}{\tau_R}

\def \be {\begin{equation} }
\def \ee {\end{equation}}
\def \bes {\begin{subequations} }
\def \ees {\end{subequations}}

\def \a {\alpha}
\def \b {\beta}

\def \d {\delta}
\def \e {\epsilon}
\def \g {\gamma}

\def \o {\omega}

\def \l {\lambda}

\def \s {\sigma}

\def \vp {\bm{p}}

\def \hp {\hat{p}}

\def \vx {\bm{x}}
\def \vr {\bm{r}}
\def \vk {\bm{k}}
\def \vv {\bm{v}}

\def \vu {\bm{u}}
\def \hp {\hat{\vp}}

\def \hk {\hat{\vk}}

\def \hO {\hat{O}}
\def \G {\Gamma}

\def \<{\langle}
\def \>{\rangle}
\def \+{\dagger}
\def \le {\left}
\def \ri {\right}

\def \tk {\tilde{k}}

\def \pd {\partial}
\def \baro {\bar{\o}}
\def \bark {\bar{k}}
\def \EHR {{\rm EHR}}

\def \in {{\rm in}}

\def \RF {{\rm R.F.\,}}

\def \qperp
\def \CC {{\cal C}}

\def \CG {{\cal G}}

\def \sG {{\cal G}}

\newcommand{\eff}{\textrm{eff}}

\def \tR {\tau_{R}}

\begin{document}
\title{Non-hydrodynamic response in QCD-like plasma}
\author[a]{Weiyao Ke}
\author[b]{Yi Yin}


\affiliation[a]{Key Laboratory of Quark and Lepton Physics (MOE) \& Institute of Particle Physics, Central China Normal University, Wuhan 430079, China}

\affiliation[b]{Quark Matter Research Center, Institute of Modern Physics, Chinese Academy of Sciences, Lanzhou, Gansu, 073000, China }

\abstract{
Quark-gluon plasma's (QGP) properties at non-hydrodynamic and non-perturbative regimes remain largely unexplored.
Here, we examine the response functions describing how a QGP-like plasma responds to initial energy-momentum disturbance in both static and Bjorken-expanding plasma at non-hydrodynamic gradient using the Boltzmann equation in the relaxation-time approximation (RTA). 
We show that the resulting response functions are remarkably similar in both static and expanding backgrounds at non-hydrodynamic gradients. 
While non-hydrodynamic response can not be described by the conventional first-order and second-order theories, its behavior is reasonably captured by the extended version of hydrodynamics proposed by us~\cite{Ke:2022tqf}. 
The potential sensitivity of the Euclidean correlator to non-hydrodynamic response is also illustrated. 
}

\emailAdd{weiyaoke@ccnu.edu.cn}
\emailAdd{yiyin@impcas.ac.cn}

\maketitle

\section{
Introduction
\label{sec:intro}
}

\begin{figure}
    \centering
    \includegraphics[width=0.75\textwidth]{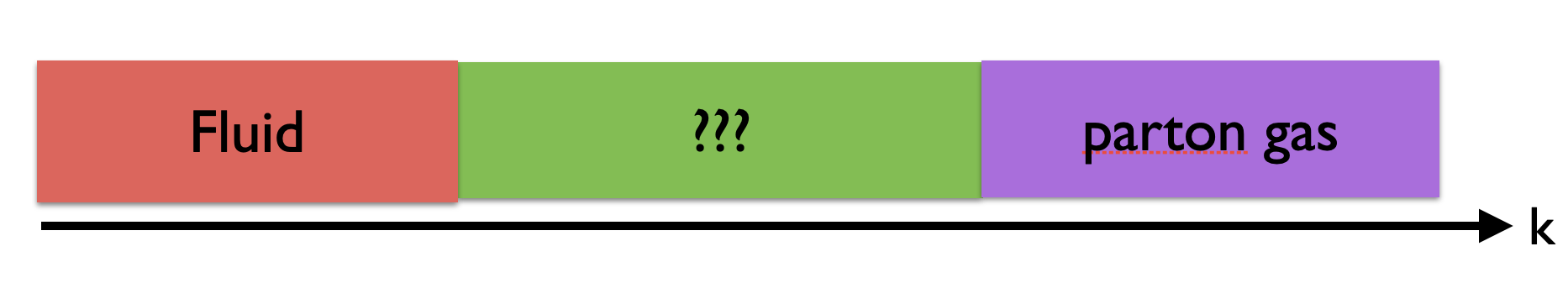}
    \caption{
        \label{fig:EHR-cartoon}
    A cartoon illustrating our current understanding of Quark-Gluon Plasma (QGP) as a function of the characteristic momentum scale ($k$). When $k$ is small, QGP behaves like a near-perfect liquid. As $k$ increases to the asymptotic free regime, QGP can be viewed as a parton gas. The properties of QGP at the intermediate scale are less known, but they can be explored by studying jet-medium interaction observables in heavy-ion collisions and by conducting the lattice QCD calculation of the Euclidean correlators, see further discussion in the text.
    }
\end{figure}

When an in-homogeneous and time-dependent disturbance is created on the quark-gluon plasma (QGP), 
its response reveals the rich structure of the medium. 
The hydrodynamic theory should describe the medium's response in long wavelength and low-frequency limits.
On the other hand, 
one could apply the perturbative QCD method to study QGP's behavior at an asymptotic large gradient. 
Apparently, exploring the intermediate regime between those two 
asymptotic limits is crucial to understanding how the properties of QCD matter evolve as scale changes. Yet, our knowledge about this non-hydrodynamic and possibly non-perturbative regime is very limited as illustrated in Fig.~\ref{fig:EHR-cartoon}. 
See Refs~\cite{DEramo:2012uzl,DEramo:2018eoy,Kurkela:2019kip,Casalderrey-Solana:2019ubu, Kurkela:2021ctp,Yang:2023dwc,Kurkela:2017xis, Heller:2016rtz} on examples of studying non-hydrodynamic properties of QGP.

By comparing the behavior of conventional hydrodynamic theories extrapolated outside the hydrodynamic regime with QGP-like models,
we may monitor how some hydrodynamic behavior disappears, and possible new features emerge as characteristic gradients gradually increase. 
Doing so can bring insight into the nature of the transition from QGP liquid to weakly-coupled parton gas.
In this work, 
we will compare the non-hydrodynamic energy-momentum response obtained from the first-order and some variants of second-order hydrodynamics with that from a particular microscopic theory, which we consider the kinetic theory under relaxation time approximation (RTA).
Despite its simplicity, the RTA kinetic theory captures a transition from the hydrodynamic regime to a regime whose response is dominated by quasi-particle excitations and has been widely used as a model for a QCD-like system. 
The study of the retarded correlator, which describes the medium's response induced by the metric perturbation using the Boltzmann equation, can be found in Refs.~\cite{Hong:2010at,Romatschke:2017vte,Kurkela:2017xis}. 
Our focus is on the response function that describes the induced energy-momentum due to the initial energy-momentum disturbance.

Another motivation of our work is to further investigate and elaborate on the notion of the extended hydrodynamic regime (EHR) proposed by us in Ref.~\cite{Ke:2022tqf}, 
This scenario corresponds to the situation that a high-frequency sound exists that governs the medium's response outside the hydrodynamic regime on the one hand, and its dispersion is not described by naively extending conventional hydrodynamic theories outside the hydrodynamic scale on the other hand. 
In Ref.~\cite{Ke:2022tqf}, we demonstrate the presence of this scenario for strongly coupled ${\cal N}=4$ super-symmetric Yang-Mills (SYM) theory and RTA kinetic theory. 
We also show that the sound dispersion can be captured by a ``deformed'' version of MIS, namely MIS*, with a suitable choice of model parameters. 
In this work, we show detailed analysis indicating that the same MIS* equations can describe not only dispersion relation but also response function outside the hydrodynamic regime.

Phenomenologically, the observables sensitive to the jet-medium interaction can help extract the medium's properties at a non-hydrodynamic scale. 
An energetic parton may be viewed as a point-like source that deposits energy and momentum density to the medium it transverses to. It would generate perturbation for a vast range of gradients, including those outside the hydrodynamic regime. 
Recently, evidence has been suggesting that properly incorporating the medium's response to those disturbances is essential in describing data~(e.g., see Refs.~\cite{KunnawalkamElayavalli:2017hxo,He:2018xjv,Cao:2020wlm,Cao:2022odi,JETSCAPE:2020uew}). 
To date, many studies are using linearized first-order and second-order hydrodynamics (or its variant) to account for the medium's response (see, for example, Refs.~\cite{Casalderrey-Solana:2004fdk,Neufeld:2008fi,Casalderrey-Solana:2020rsj}). 
We have analyzed the energy-momentum response for a Bjorken-expanding background. 
This allows us to compare the response functions for the static and expanding background and discuss the importance of non-hydrodynamic response in the context of jet-medium interaction.

Theoretically, one may extract the medium's properties from the Euclidean correlator, which can be studied from the first-principle lattice QCD calculations. 
While the Euclidean correlators are notoriously insensitive to the infrared structure of the real-time response (see ref.~\cite{Moore:2010bu} for a review), it may be suitable for extracting response at a gradient of the order $\pi T$. 
We present perhaps the first illustrative analysis that suggests the sensitivity of the Euclidean correlator to the non-hydrodynamic response. 
Assuming the EHR scenario, we also explore the lattice's potential to extract the medium's properties in EHR.

This paper is organized as follows.
In sec.~\ref{sec:R-C-relation},
we derive a general relation between the response function describing the response induced by initial disturbance and the retarded correlator describing the medium's reaction to the metric perturbation, see eq.~\eqref{G-relation} (or eq.~\eqref{G-relation-o}). 
While it directly follows from the standard linear response theory, it seems new in the literature. 
We then compute the energy-momentum response function in a static background from RTA kinetic theory and hydrodynamic theories in sec.~\ref{sec:kin-G} and~\ref{sec:hydro-G} respectively. 
In sec.~\ref{sec:excitation}, 
we analyze the excitations that dominate the response for the RTA kinetic theory. 
In doing so, 
we review the concept of EHR and discuss its generality. 
Then we review the construction of MIS* in sec.~\ref{sec:MIS-STAR}.
Sec.~\ref{sec:static-com} compares kinetic response and that from different hydrodynamic theories. 
In sec.~\ref{sec:B-formalism}, we set up the formalism to calculate similar response functions in a Bjorken expanding background and present results in \ref{sec:Bjorken-result}. 
The sensitivity of the Euclidean correlator to non-hydrodynamic response is discussed and illustrated in sec.~\ref{sec:GE}.
Finally, we summarize in sec.~\ref{sec:outlook}.

Throughout this paper, we use the most plus metric $\eta^{\mu\nu}=(-1,1,1,1)$.

\section{Response in a static background
\label{sec:static-response}
}

\subsection{Response function and retarded correlator
\label{sec:R-C-relation}
}

We want to investigate the behavior of a thermal medium when it is in equilibrium and then undergoes an infinitesimally small perturbation. 
For definiteness, we look at a set of observables such as energy, and charge densities and collectively denoted the corresponding operator by $\hat{O}^{I}$ where $I,J$ are the labels. 
In response to the perturbation, the expectation value $O=\langle \hat{O}\rangle$ will deviate from its equilibrium value $\langle \hat{O}\rangle_{{\rm eq} }$, i.e. $\delta O=O-\langle \hat{O}\rangle_{{\rm eq} } \neq 0$.

The present work primarily considers the response induced by the out-of-equilibrium deviation $\delta O^{I}_{\in}(\vx)\equiv \delta O^{I}(t=0,\vx)$ at some initial time, say $t=0$.
To describe such a response, we define response function (\RF) $\sG$ as
\begin{align}
\label{Response-def}
    \delta O^{I}(t,\vx)=\int_{\vx'}\sG^{I}_{\,J}(t,\vx-\vx')\,\delta O^{J}_{\in}(\vx') \, 
\end{align}
where $\int_{\vx}\equiv\int d^{3}\vx $ and the summation over the dummy index is understood.
We shall assume that $\sG$ satisfies the following initial condition:
\begin{align}
\label{response-init}
    \sG^{I}_{\, J}(t=0,\vx)=\delta^{I}_{\, J}\,\delta^{3}(\vx)\, ,
\end{align}
meaning we have assumed that when $I\neq J$, a non-zero $\d O_{I}$ will not cause a non-zero $\d O_{J}$ at the beginning. 
The discussion below can be easily generalized when this assumption is relaxed.

In parallel, 
the retarded correlator
\begin{align}
\label{G-def}
    G_{R}^{IJ}(t, t';\vx, \vx')\equiv -i \langle\, \le[\hO^{I}(t,\vx),\hO^{J}(t',\vx')\ri] \rangle\, \theta(t-t')
\end{align}
has been widely considered in literature. 
This correlator describes the response induced by the source $\psi_{I}(t,\vx)$ that is conjugate to $\hat{O}^{I}$ asuming that the perturbed Hamiltonian takes the form $\Delta \hat{H}(t)=-\int_{\vx}\, \hat{O}^{J}(t,\vx)\psi_{J}(t,\vx) $.

We can establish the relation between \RF and the retarded correlator. 
The key observation is that $\sG$ is entirely determined by the medium's properties and hence does not depend on how the initial deviation $\delta O_{\in}$ is created. 
We may generate $\delta O^{J}_{\in}$ by turning on the source $\psi_{J}$ at $t=-\infty$ and switch it off at $t=0$, i.e.,
\begin{align}
\label{source-0}
    \psi_{J}(t,\vx)=\theta(-t)\, e^{-\e t}\psi_{J,\in}(\vx)\, ,
\end{align}
where an infinitesimal value $\e$ is introduced to ensure that the source vanishes at $t=-\infty$. 
Using the linear response theory (e.g., ref~\cite{Petreczky:2005nh}), we then have, after taking spatial Fourier transform (assuming the translational invariance in space and time), that
\begin{align}
\label{lin-G}
    \delta O^{I}(t,\vk) = \int^{\infty}_{-\infty}\,dt'\,G^{IJ}_{R}(t-t',\vk)\, \psi_{J}(t',\vk)\, .
\end{align}
Particularly, evaluating eq.~\eqref{lin-G} at $t=0$ with $\psi_{J}$ specified by eq.~\eqref{source-0} yields
\begin{align}
\label{chi}
    \delta O^{I}_{\in}(\vk)=\le(\int^{0}_{-\infty} dt'\, G_{R}^{IJ}(-t',\vk)e^{\e t'}\ri)\, \psi_{J,\in}(\vk)
    =\chi^{I J}(\vk)\, \psi_{J,\in}(\vk)\, .
\end{align}
In other words, the induced $\delta O^{I}_{\in}$ resulting from turning on a time-independent source $\psi_{J,\in}$ from $t=-\infty$ to $t=0$ is determined by the static susceptibility.
\begin{align}
\label{sus}
    \chi^{IJ}(\vk)=\int^{\infty}_{0} dt\, G^{IJ}_{R}(t,\vk)=
    G^{IJ}_{R}(\omega=0,\vk)\,. 
\end{align}

Now, we take the time derivative of eq.~\eqref{lin-G} w.r.t time $t$. 
After using $\pd_{t}G_{R}(t-t')=-\pd_{t'}G_{R}(t-t')$ and performing the integration over $t'$ by parts,
we obtain (see also eq.~2.7 of Ref.~\cite{Petreczky:2005nh})
\begin{align}
\label{O-source}
    \pd_{t}\delta O^{I}(t,\vk)=-G_{R}^{IJ}(t,\vk)\psi_{J,\in}(\vk)\, . 
\end{align}
Assuming $\chi^{IJ}$ is invertible, we can rewrite \eqref{O-source} as
\begin{align}
\label{dOdt-GR}
    \pd_{t}\delta O^{I}(t,\vk)&=
    -G_{R}^{IJ'}(t,\vk)\, (\chi^{-1}(\vk))_{J'J}\delta O^{J}_{0}(\vk)\, . 
\end{align}
If we take the derivative of eq.~\eqref{Response-def} and compare it with eq.~\eqref{dOdt-GR}, we find the relation between the response function and retarded correlator that we are after:
\begin{align}
\label{G-relation}
    \pd_{t}\sG^{I}_{J}(t,\vk)=-G_{R}^{IJ'}(t,\vk)\, (\chi^{-1}_{R}(\vk))_{J'J}\, .
\end{align}
Further performing the temporal Fourier transformation, we have
\begin{align}
\label{G-relation-o}
    i\o\sG^{I}_{J}(\o,\vk)+\d^{I}_{J}=G_{R}^{IJ'}(\o,\vk)\, (\chi^{-1}(\vk))_{J'J}
\end{align}
where we have used the property of "one-sided" Fourier transform
\begin{align}
\label{dt-FT}
    \int^{\infty}_{0} dt\, e^{i\o t}\,\pd_{t}\,g(t)= -i \o g(\o,\vk)-g_{\in}\, . 
\end{align}
for any function $g(t)$ satisfying $g(t=0)=g_{\in}$ and the initial condition for $\sG$~\eqref{response-init} in the spatial Fourier space :
\begin{align}
\label{response-init-k}
    \sG^{I}_{J}(t=0,\vk)=\delta^{I}_{J}\, .
\end{align}

Eqs.~\eqref{G-relation} and \eqref{G-relation-o} are the main results of this section. 
They make it clear that one can determine the \RF from the knowledge of the retarded correlator. In other words, 
how the medium reacts to the external source also tells us how it would respond to the initial perturbation. 
Moreover, the relation~\eqref{G-relation-o} reveals that the analytical structure of \RF is the same as that of the retarded correlator. For instance, the pole of the correlator in the complex frequency plane should be the pole of \RF at exactly the same location.

In the upcoming sections, we will discuss \RF of the energy-momentum tensor (EMT):
\begin{align}
 \delta T^{\mu\nu}(t,\vk)=\sG^{\mu\nu}_{\,\,\,\,\,\a\b}(t,\vk)\, \delta T^{\a\b}_{\in}(\vk)\, . 
\end{align}
Our focus will be on a charge-neutral and conformal fluid. 
In this situation, there are only three independent components in $\sG^{\mu\nu}_{\,\,\a\b}$. Among them, one is associated with the behavior of shear stress. We will concentrate on the remaining two, which we choose to be
\footnote{
It is well-known that by implementing Ward identity and rotational symmetry, the retarded correlator has only three independent components for a conformal system at finite temperature~\cite{Kovtun:2005ev}. Among them is one representative component of shear stress correlator, while the other two can be identified as $G^{00,00}_{R}\sim \langle \d T^{00}\d T^{00}\rangle$ and $G^{0x,0x}_{R}\sim\langle \d T^{0x}\d T^{0x}\rangle$.
The components of \RF can be related to those of retarded correlator from the relation~\eqref{G-relation-o}.
}
\begin{align}
\label{GLT}
    \sG_{L}(t,k)\equiv  \sG^{00}_{\,\,\,\,\,00}(t,k)
    \qquad
    \sG_{T}(t,k)\equiv \sG^{0x}_{\,\,\,\,\,0x}(t,k)\, 
\end{align}
where without losing generality, we take the Fourier momentum $\vk$ along $z$-direction.

\subsection{Kinetic theory
\label{sec:kin-G}
}

The response function has to be computed from some microscopic theories to which we take the kinetic equation under relaxation time approximation (RTA)
\begin{align}
\label{RTA}
  p^{\mu}\pd_{\mu}f-\Gamma^{\l}_{\a\sigma}p^{\a}p^{\sigma}\frac{\pd f}{\pd p^{\l}} = -\frac{u\cdot p}{\tau_{R}}\, \le(
f - f_{{\rm eq}}
\ri)\, .
\end{align}
Here,  $f(t,\vx,\vp)$ is the distribution of relativistic particles and $\G^{\l}_{\a\b}$ is the metric connection. 
The relaxation time $\tau_{R}$ sets the timescale at which $f$ approaches the local equilibrium
\begin{align}
\label{feq}
f_{{\rm eq}}\equiv e^{\beta p^{\mu} u_{\mu}}\, . 
\end{align}
The local fluid velocity, $u^{\mu}$ together with energy density $\e$, are determined by the matching condition
\begin{align}
\label{T-f}
    T^{\mu\nu}=\int_{\vp}\, \frac{1}{p}\,p^{\mu}\,p^{\nu}\, f\, ,
    \qquad
T^{\mu\nu}u_\nu = \epsilon\,u^\mu\, . 
\end{align}
where $\int_{\vp}\equiv \int d^{3}\vp/(2\pi)^{3}$, 
and for ultra-relativistic particle, we have $p=|\vp|$. 
With $\e$ given, the effective temperature $T=1/\beta$ in eq.~\eqref{feq} is determined by assuming the equation of state for the relativistic Boltzmann gas, i.e.,
$\e(T)= N_{0} (3 T^{4})/\pi^{2}$ where $N_{0}$ counts the quantum state and species of particles under consideration, although the results shown in the present paper do not depend on the value of $N_{0}$.
For the system to be conformal system, we require $\tau_{R}$ scales with $\e$
\begin{align}
    \label{tau-scale}
    \tau_{R}\sim (\e)^{-1/4}\, . 
\end{align}

We shall study \RF~\eqref{GLT} describing the induced energy density and fluid velocity around the equilibrium background
\begin{align}
\label{hydro-var-lin}
    \e=\e_{0}+\delta\e\, , 
    \qquad
    u^{\mu}=(1,{\bm 0})+ (0,\delta \vu)\, ,
\end{align}
where we use subscript $0$ to denote the background value of thermodynamic and fluid variables such as $\e, u^{\mu}$.
Using 
\begin{align}
\label{delta-f-eq}
\delta f_{\textrm{eq}} = 
e^{-\beta_{0}p^{0}}
\left(
-p_{0}\delta\beta+\beta_{0}\, {\vp}\cdot \delta {\bm u}
\right)\,  
\end{align}
we have
\begin{align}
\label{kin-lin}
    \le(\pd_{t}+\vv\cdot {\bm \pd} +\tau^{-1}_{R}\ri)\, \delta f 
    = \beta p_{0}\,f_{0}\,\tau^{-1}_{R}\le[
    -\frac{\delta \beta}{\beta}+\delta\vu\cdot\vv
    \ri]\, .
\end{align}
where we have introduced the short-handed notation $f_{0}\equiv e^{-\beta_{0}p^{0}}$ for the background equilibrium distribution and $\vv=\hat{p}$ is the single particle velocity. 
We shall solve eq.~\eqref{kin-lin} under the initial condition that the distribution is in local equilibrium:
\begin{align}
\label{fI}
    \delta\, f_{I}=\delta\,e^{\beta(p\cdot u)}=\beta_0 p^0\, f_{0}\le(\frac{-\delta\beta_{I}}{\beta_{0}}+\delta \vu_{I}\cdot \vv \ri)\, ,
\end{align}
where $\delta \vu_{I}, \delta \beta_{I}$ denotes the value of $\delta\vu,\delta \beta$ at $t=0$. 
The solution to eq.~\eqref{kin-lin} in Fourier space then reads
\begin{align}
\label{delta-f-sol}
\delta f(\baro, \bark; \vv) = \frac{i\,\beta_{0}p_{0}f_{0}\tau_R^{-1}}{(\o-\vv\cdot\bar{\vk} + i\tau_R^{-1})}\, 
\le[\left(-\frac{\delta\beta(\baro, \bar{\vk})}{\beta_0}+\delta \vu(\baro, \bar{\vk})\cdot\vv\right)+\frac{\delta f_I(\bar{\vk};\vv)}{\beta_{0}p_{0}f_{0}\tau_R^{-1}}\ri]\, .
\end{align}
where  $\baro=\omega\tau_{R}$ and $\bark=k\tau_{R}$. 
Plug in the Fourier transform of the initial condition, and further use the relation $\b^{2}\d \epsilon=-c_{V}\delta\beta$ with $c_V=\pd\epsilon/\pd T$ being the specific heat, we arrive at
\begin{align}
\delta f(\baro,\bark) = \frac{i\,\beta_{0}p_{0}f_{0}}{(\baro-\vv\cdot\bar{\vk} + i)}\, 
\le[\frac{\beta_{0}\delta\epsilon(\baro, \bar{\vk})}{c_{V}}+\delta \vu(\baro, \bar{\vk})\cdot\vv+\left(\frac{\beta_{0}\delta\epsilon_{I}(\bar{\vk})}{c_{V}}+\delta \vu_{I}(\bar{\vk})\cdot\vv\right)\,\tau_{R}\ri]\, .
\end{align}
Note that the mass dimension of $\delta \vu(\baro,\bark)$ is $-4$ while that of $\delta \vu_{I}(\bark)$ is $-3$.

We are now ready to determine the \RF. 
For the response in the longitudinal channel, we consider non-zero $\delta \e, \delta u^{z}$ where, as we mentioned earlier, we shall take $\vk$ along the $z$-direction. 
We are interested in computing energy-energy response function $\sG_{L}$~\eqref{GLT} so we use the initial condition $(\delta \e, \delta u^{z})=(\delta \e_{\in}, 0)$ at $t=0$ in accordance with eq.~\eqref{response-init}) 
Substituting eq.~\eqref{delta-f-sol} into the linearized version of eq.~\eqref{T-f}, 
\begin{align}
\label{delta-T-f}
    \delta T^{\mu\nu}=\int_{\vp} \frac{p^{\mu}p^{\nu}}{p}\, \delta f\, ,
\end{align}
and use the relation
\begin{align}
\label{e-match}
\delta \e= \delta T^{00}, 
\qquad
\delta T^{0i}=w_{0}\delta u^{i}
\end{align}
where $w=\e+P$ is the enthalpy density, 
we find
\bes
\label{delta-ep-I}
\begin{align}
&\,    \delta \e(\o,\vk) = C_{0}(\o,\vk)(\delta \epsilon(\o,\vk)+\delta\epsilon_{\in}(\vk)\tau_{R}) + \frac{1}{c^{2}_{s}}\,C_{1}(\o,\vk)\,\delta T^{0z}(\o,\vk)\, , 
\\    
&\,  \delta T^{0z}(\o,\vk)=C_{1}(\o,\vk)(\delta \epsilon(\o,\vk)+\epsilon_{\in}(\vk)\tau_{R}) + \frac{1}{c^{2}_{s}}\,C_{2}(\o,\vk)\,\delta T^{0z}(\o,\vk)\, .
\end{align}
\ees
Here, we have used the expression for $c_{V}$
\begin{align}
    c_{V}= \int_{\vp}p\frac{\pd f_{0}}{\pd T }= \beta^{2}\int \frac{d p}{2\pi^{2}}\, p^{2}f_{0}\, 
\end{align}
which leads to the following useful expression
\bes
\label{d-hydro-f}
\begin{align}
&\,    \int\, \frac{dp}{2\pi^{2}}\, \beta_{0} p^{2}f_{0}(-\frac{\d \beta}{\beta_{0}})\, (\ldots)
    = \beta^{-2}_{0}\, c_{V}(-\delta \beta)\int \frac{d\Omega}{4\pi}\, (\ldots)= \delta \epsilon \int \frac{d\Omega}{4\pi}\, (\ldots)
\\
&\,    \int\,\frac{dp}{2\pi^{2}}\, \beta_{0} p^{2}f_{0}\,\delta u^{i}\, (\ldots)
    = \frac{\delta T^{0i}}{w_{0}}\beta^{-2}_{0}\, c_{V}\int \frac{d\Omega}{4\pi}\, (\ldots)= \frac{\delta T^{0i}}{c^{2}_{s}} \,\int \frac{d\Omega}{4\pi}\, (\ldots) \, ,
\end{align}
\ees
Here, $d\Omega=d\phi\, d\cos\theta$ denotes the integration over the solid angle in the phase space, and $(\ldots)$ refer to any functions which only depend on $\hat{p}=(\sin\theta\cos\phi,\sin\theta\cos\phi,\cos\theta)$. 
The sound velocity is denoted by $c^{2}_{s}$ and obeys the relation $c_{V}T/w = 1/c^{2}_{s}$. 
In eq.~\eqref{delta-ep-I}, functions $C_{n}$ are defined by
\begin{align}
    C_{n}\equiv \frac{1}{2}\int^{1}_{-1}\,d\cos\theta \, \frac{i\cos^{n}\theta}{\baro-\bark\,\cos\theta+i}\, .
\end{align}
Solving eq.~\eqref{delta-ep-I} for $\delta \e, \delta \vu$ and using $c^{2}_{s}=1/3$ yields $\delta \e$ induced by the initial energy disturbance
\begin{align}
&\,    \delta \epsilon= \frac{-\le(-3 C^{2}_{1}+3 C_{0}C_{2}-C_{0}\ri) \e_{\in}(\vk)\tau_{R}}{-3 C^{2}_{1}+3C_{0}C_{2}-C_{0}-3C_{2}+1}\, , 
\end{align}
from which we read the response function
\begin{align}
\label{GL-kin}
\, \sG_{L}(\o,k)
&=\frac{-\le[6 \bark + \le(3(i+\baro)+i\bark^{2}\ri)L(\baro,\bark)\ri]\,\tR}{ 2\bark(\bark^2+3i\baro)+i\le(\bark^2+3\baro(i+\baro)\,\ri)L(\baro,\bark)}\, .
\end{align}
Here, for convenience,  we have defined
\begin{align}
\label{L-fun}
    L(\o,k)= \ln\le(\frac{\baro-\bark+i}{\baro+\bark+i}\ri)\, .
\end{align}
and used the relation among $C_{n}$s. 
\bes
\label{C12}
\begin{align}
\label{C0}
    C_{0}&=\frac{-i}{2 \bark}\, L(\baro,\bark)\, , 
\qquad
    C_{1}=  
    \frac{-i}{ \bark}\,(1-(1-i\baro)C_{0})
  \, . 
    \\
    C_{2}&=\frac{1}{\bark}\,(i+\baro)\, C_{1}
    = \frac{(1-i\baro)}{\bark^{2}}\,
    (1-(1-i\baro)C_{0}) \, . 
\end{align}
\ees

Similarly, we find that in the transverse channel
\begin{align}
\label{GT-kin}    
\,   \sG_{T}(\o,k)
&=\frac{3i \le[2\bark(\baro+i)-\le(\bark^{2}+(1-i\baro)^{2}\ri)L(\baro,\bark)\ri]\,\tR}{2\bark(3+2\bark^{2}-3i\baro)+3i\le(\bark^{2}+(1-i\baro)^{2}\ri)L(\baro,\bark)}\, . 
\end{align}

The expression for the RTA retarded correlators has been obtained in Ref.~\cite{Romatschke:2015gic}, see also Ref.~\cite{Kurkela:2017xis}. 
We copy the relevant expression in terms of $C_{0,1,2}$ below (up to a contact term) in our notation: 
\begin{align}
&\,    G^{00,00}_{R}(\o,k)=3w_{0}\,\frac{i \bark C_{1} }{-3 C^{2}_{1}+3C_{0}C_{2}-C_{0}-3C_{2}+1}\, , 
    \\
&\,    G^{0x,0x}_{R}(\o,k)=
-3 w_{0}\,\frac{i\baro(C_{0}-C_{2})}{2-3(C_{0}-C_{2})}\, . 
\end{align}
Using the expressions \eqref{GL-kin} and \eqref{GT-kin}, one can verify the relation~\eqref{G-relation-o} between the response function and the retarded correlator.

\subsection{Hydrodynamic response function
\label{sec:hydro-G}
}

The primary goal of this paper is to investigate the response function's behavior beyond the hydrodynamic regime. 
To benchmark hydrodynamic behavior, we now calculate $\sG_{L}$ and $\sG_{T}$ within hydrodynamics and/or its cousin theory,  Muller, Israel and Stewart (MIS) theory. 
In the following section, we will compare these results with the RTA response.

We begin with the decomposition of EMT
\begin{align}
\label{pi-def}
    T^{\mu\nu} =\e u^{\mu}u^{\nu}+P\,\Delta^{\mu\nu} + \pi^{\mu\nu}
\end{align}
where $p$ denotes the pressure, and the projector is given by
\begin{align}
    \Delta^{\mu\nu}= \eta^{\mu\nu}+u^{\mu}u^{\nu}\, , 
\end{align}
and $\pi^{\mu\nu}$ arise from the gradient correction to $T^{\mu\nu}$.
For a conformal fluid, $P=\e/3$ and $\pi^{\mu\nu}$ is traceless.
One may expand $\pi^{\mu\nu}$ in gradient in the long time and wavelength limit. 
To the first-order gradient, 
$\pi^{\mu\nu}$ takes the form
\begin{align}
\label{pi-vis}
    \pi^{\mu\nu}_{(1)}=-2\,\eta\, \sigma^{\mu\nu}
\end{align}
where $\eta$ is the shear viscosity, and the shear stress is given by 
\begin{align}
\label{shear}
    \sigma^{\mu\nu}= \nabla^{\langle\mu}\, u^{\nu\rangle}\, .
\end{align}
Here $\nabla^{\mu}$ denotes the covariant derivative. 
Angular brackets around a pair of Lorentz indices, $\,^{<\mu\nu>}$, mean that the indices are to be symmetrized, space-projected, and trace-subtracted, i.e.:
\begin{align}
    A^{<\mu\nu>}= \frac{1}{2}\Delta^{\mu\alpha}\Delta^{\mu\beta}\le(A_{\a\b}+A_{\b\a}\ri)-\frac{1}{3}\Delta^{\mu\nu}\Delta^{\a\b}\, A_{\a\b}\, . 
\end{align}
As usual, 
we refer to the conservation equation
\begin{align}
\label{T-conser}
    \nabla_{\mu}T^{\mu\nu}=0
\end{align}
supplemented by the relation $\pi^{\mu\nu}=\pi^{\mu\nu}_{(1)}$~\eqref{pi-vis} as the first-order hydrodynamics for a charge-neutral system. 
One may include contributions from higher-order gradient terms~\cite{Baier:2007ix,Bhattacharyya:2007vjd,Grozdanov:2015kqa}.
Various versions of higher-order theories have shared a feature that $\pi^{\mu\nu}$ is promoted as an independent dynamical degree of freedom.
For the theory developed by Muller, Israel and Stewart (MIS), 
$\pi^{\mu\nu}$ obeys a simple relaxation equation:
\begin{align}
\label{MIS}
    D\pi^{\mu\nu}=- \frac{1}{\tau_{\pi}}\le(\pi^{\mu\nu}-\pi^{\mu\nu}_{(1)}\ri)\, , 
\end{align}
where $D\equiv u\cdot \nabla$ is the time-derivative in the fluid rest-frame and $\tau_{\pi}$ gives the relaxation time scale under which $\pi^{\mu\nu}$ approaches $\pi^{\mu\nu}_{(1)}$ .

We first consider the response function in the longitudinal channel obtained from linearizing hydrodynamic theories. 
As we did earlier, we take $k$ along $z$-direction. 
Therefore the relevant dynamical fields are $\delta\e, \delta u^{z}, \pi^{zz}$ where shall count 
$\pi^{ij}$ as ${\cal O}(\d)$. 
To compute $\sG_{L}$~\eqref{GLT},  it is sufficient to consider the initial condition $(\d\e,\d u^{z}, \pi^{zz})\Big|_{t=0}=(\delta\e_{\in},0,0)$.

By linearizing conservation equation~\eqref{T-conser} and evolution equation for $\pi^{\mu\nu}$~\eqref{MIS} and the performing the Fourier transform, we have
\bes
\label{hydro-lin-h0}
\begin{align}
\label{e-lin}
-i\o \, \frac{\delta\e}{w_{0}}&=  -i k\delta u^{z}+\frac{\delta\e_{\in}}{w_{0}} \, , 
    \\
    \label{uz-lin}
     -i\o\,\delta u^{z}&=-i k\le(\, c^{2}_{s}\frac{\delta \epsilon}{w_{0}}+\, \frac{\pi^{zz}}{w_{0}}\ri)\, ,
     \\
     \label{pizz}
      - i\o \frac{\pi^{zz}}{w_{0}} &= - \frac{1}{\tau_{R}}\le(\frac{\pi^{zz}}{w_{0}}+i k\,\frac{4}{3}\,\nu_{0}\,\delta u_{z} \ri)\, , 
\end{align}
\ees
where specific viscosity is given by $\nu=\eta/w$. 
To obtain eq.~\eqref{e-lin}, we have used eq.~\eqref{dt-FT}. 
Now, we solve eq.~\eqref{pizz} to find
\begin{align}
\label{pizz-sol}
    \frac{\pi^{zz}}{w_{0}}= -\nu(\o)\,i k \delta u^{z}\,  ,
\end{align}
where we have defined frequency-dependent longitudinal momentum diffusive coefficient
\begin{align}
\label{nu-L-MIS}
    \nu(\o)=\frac{4}{3} \frac{\nu_{0}}{1- i \o\tau_{\pi}}\,\, ,
\end{align}
Substituting eq.~\eqref{pizz-sol} into eq.~\eqref{uz-lin} and using eq.~\eqref{e-lin} yields
\begin{align}
\label{hydro-sol}
    \d \e(\o,k) &= \frac{1}{R(\o,k)}\,
    \le(i \o -\nu(\o)k^{2}\ri)\,\delta \e_{\in}(\o,k)\, , 
\end{align}
where 
\begin{align}
\label{R-def}
    R(\o,k)= \o^{2}-c^{2}_{s}k^{2}+ i \nu(\o)\o k^{2}\, . 
\end{align}
We then arrived at the expression for the longitudinal response function
\begin{align}
\label{GL-hydro}
    {\cal G}_{L}(\o,k)&= \frac{i\o-\nu(\o)k^{2}}{R(\o,k)}\, .
\end{align}

In parallel, 
we find
\begin{align}
\label{GT-hydro}
    {\cal G}_{T}(\o,k)&= \frac{i}{\o+i \nu_{T}(\o)\ k^{2}}\, , 
\end{align}
where frequency-dependent transverse momentum diffusive coefficient reads (c.f. eq~\eqref{nu-L-MIS})
\begin{align}
\label{nu-T-MIS}
\nu_{T}(\o)= \frac{\nu_{0}}{1- i \o\tau_{\pi}}\, . 
\end{align}
It is easy to verify that the above hydrodynamic expressions can be matched to the small $\o, k$ behavior of RTA response function~\eqref{GL-kin}, \eqref{GT-kin} with
\begin{align}
\label{H-para}
c^{2}_{s}=\frac{1}{3}\, , 
\qquad
\nu_{0}=\frac{1}{5}\, \tR
\qquad
\tau_{\pi}=\tR\, 
\end{align}

Before closing, we recall the related EMT correlators in the hydrodynamics regime (see e.g., Refs.~\cite{Hong:2010at,Romatschke:2017vte}) is given by (up to the $\o,k$-independent contact terms) 
\begin{align}
\label{GR-hydro}
G^{00,00}_{R}(\o,k)=\frac{w_{0}k^{2}}{R(\o,k)}\, , 
\qquad
G^{0x,0x}_{R}(\o,k)=\frac{-iw_{0}\,\nu_{T}(\o)k^{2}}{\o+i \nu_{T}(\o)k^{2}}\
\end{align}
It is then straightforward to confirm the relation between the response function and the retarded correlator~\eqref{G-relation-o}.

\section{Extended hydrodynamic regime
\label{sec:EHR}
}

\subsection{Excitations
\label{sec:excitation}
}

The behavior of the \RF is determined by the properties of the medium's excitations that can be recognized by studying its analytic structure (pole, branch-cut) in the complex $\o$-plane. 
In physical terms, poles indicate the collective modes and the branch-cuts are typically associated with quasi-particle excitations.
In the time domain, 
\begin{align}
\label{G-t}
    \CG( \Delta t, k) &= \int^\infty_{-\infty} \frac{d\omega}{2\pi}\, \CG(\omega, k) e^{-i\omega \Delta t}
\end{align}
we can close the contour of the integration over $\o$ in the lower half-plane to explicitly show the contributions of the pole and branch cut, 
\begin{align}
\label{G-decom}
\CG( \Delta t, k) &= -i \sum_{n} \mathcal{R}_n(k) e^{-i\omega_n \Delta t} + \sum_{m}\oint_{\rm \mathcal{C}_{m}} \frac{d\omega}{2\pi} \CG(\omega, k) e^{-i\omega \Delta t}\, ,
\end{align}
where all the poles and branch cuts are labeled by $n$ and $m$, respectively. 
Here, $\mathcal{R}_n(k)$ is the residue of the $\CG(\omega, k)$ at pole $\omega=\omega_n(k)$. $\mathcal{C}_m$ schematically denote the contour around the branch-cut $m$.

Specifically, the RTA response function has poles given by the zeros of the denominator in eqs.~\eqref{GL-kin},\eqref{GT-kin}:
\bes
\begin{align}
\label{sound-kin}
&\,    2\bark\,(\bark^2+3i\baro)+i\le(\bark^2+3\baro(i+\baro)L(\baro,\bark)\ri)=0\, , 
\\
\label{shear-kin}
&\, 
2\bark\,(3+2\bark-3i\baro)+3i\le(\bark^{2}+(1-i\baro)^{2}\ri)L(\baro,\bark)=0\, ,
\end{align}
\ees
for the longitudinal and transverse channels, respectively. 
In the longitudinal channel~\eqref{sound-kin}, 
the solutions present a pair of propagating modes (sound-like modes) $\o_{L}(k)=\pm v(k) k-i \Gamma(k)$ where $v$ is the phase velocity and $\Gamma$ is the attenuation rate.
In the transverse sector~\eqref{shear-kin}, 
the solution is purely imaginary $\o_{T}=-i \Gamma_{T}(k)$.
In the regime where $k$ is small, 
RTA collective modes are usual hydrodynamic sound and shear modes (see also Fig.~\ref{fig:dispersion} below). 
Nevertheless, as first noticed in Ref.~\cite{Romatschke:2015gic}, collective modes in RTA continue to exist until $\bark=\bar{k}_{O}$ with $\bar{k}_{O}=4.52$ for the longitudinal channel and $\bar{k}_{O}=3\pi/4=2.37$ for the transverse channel. 
We refer to those modes outside the hydrodynamic regime as high-frequency sound and shear modes for the clarity of presentation and for the reasons that will be revealed later in Sec.~\ref{sec:EHR-gen}. 
\footnote{
It is fun to note that the high-frequency sound mode may be viewed as analogous of
Tiktaalik in the context of biological evolution. 
According to Wikipedia (\href{}{https://en.wikipedia.org/wiki/Tiktaalik}), 
``its fins have thin ray bones for paddling like most fish, but they also have sturdy interior bones that would have allowed Tiktaalik to prop itself up in shallow water and use its limbs for support as most four-legged animals do. Those fins and other mixed characteristics mark Tiktaalik as a crucial transition fossil, a link in evolution from swimming fish to four-legged vertebrates.''
Similarly, the high-frequency sound mode propagates as a sound mode but with a phase velocity closer to the velocity of parton than the ordinary sound. It might be the crucial link in the evolution of excitations in a QGP-like plasma from hydrodynamic modes to the weakly-coupled quasi-particles. 
}

Besides the poles, $\sG_{L,T}$ have a logarithmic branch cut stretching from $\bar{\omega}=-\bar{k}-i$ to $\bar{\omega}=\bar{k}-i$ (see the expression of $L$~\eqref{L-fun}). 
They represent quasi-particle excitations through the effect akin to Landau-damping; see Sec.~IIB of Ref.~\cite{Kurkela:2017xis} for an intuitive elaboration on their origin. 
For $\bar{k}>\bark_{O}$, the response of the medium is determined solely by quasi-particle excitations. At the same time, hydrodynamic modes govern for $\bar{k}< k_{H}\, \tau_{R}$, with $k_{H}$ denoting the boundary of the hydrodynamic regime.
The question of interest, therefore, is, what is the relative importance of "high-frequency" sound and shear modes and quasi-particle excitations in the intermediate regime $k_{H}<\bar{k}<\bark_{O}$?

\begin{figure}
    \centering
    \includegraphics[width=0.45\textwidth]{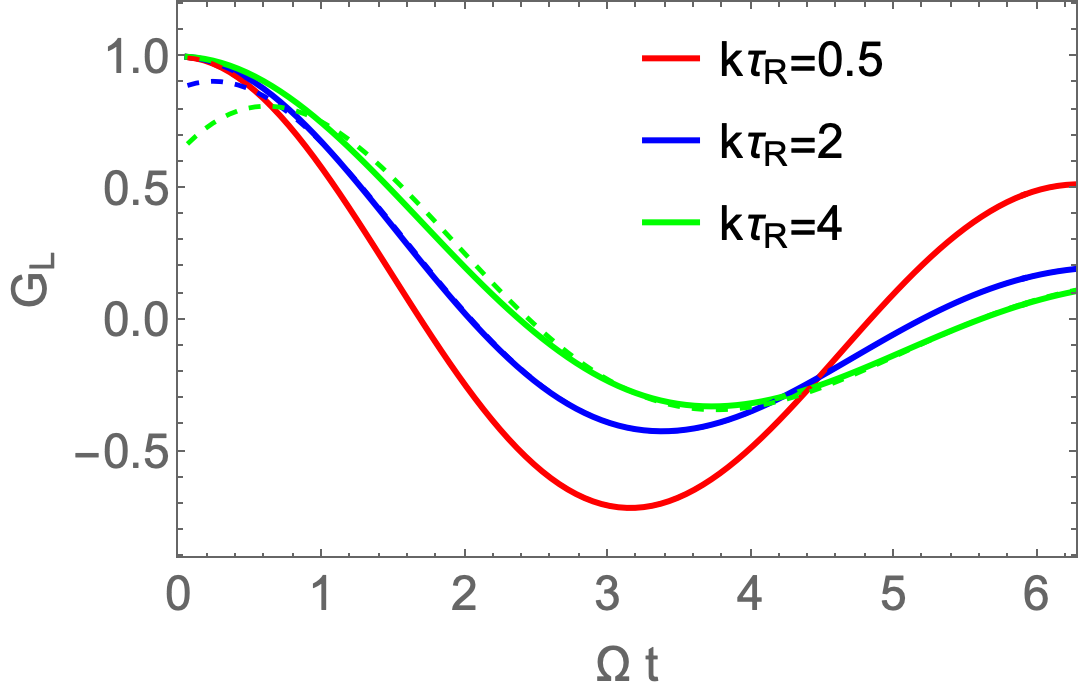}
    \includegraphics[width=0.45\textwidth]{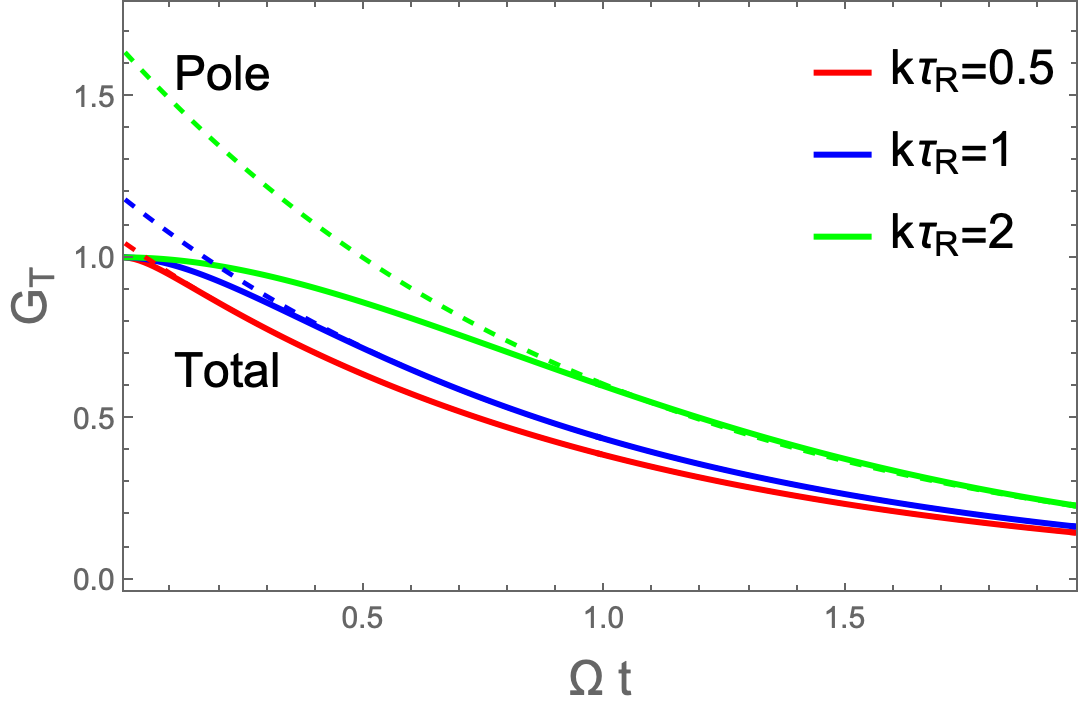}    
    \caption{
        \label{fig:RTA-decom}
    RTA response function in the time domain (solid curves) at three different values of $\bark=k\tau_{R}$. 
    They are compared with the contribution from the collective modes $\omega_{L}(k), \omega_{T}(k)$ (dashed curves), which are conventional sound and shear modes in hydrodynamic regime (see eq.~\eqref{G-decom} and the text).
    In the extended hydrodynamic regime for RTA kinetic theory where $\bark \geq 1$, those modes are referred to as high-frequency sound and shear mode. 
     (Left): the longitudinal channel where we have rescaled the time by $\Omega=|{\rm Re}\Omega_{L}(k)|$ 
     (Right): the transverse channel where we have rescaled the time by $\Omega=|{\rm Im}\Omega_{T}(k)|$.  
    }
\end{figure}

%
%
\begin{figure}
    \includegraphics[width=.32\textwidth]{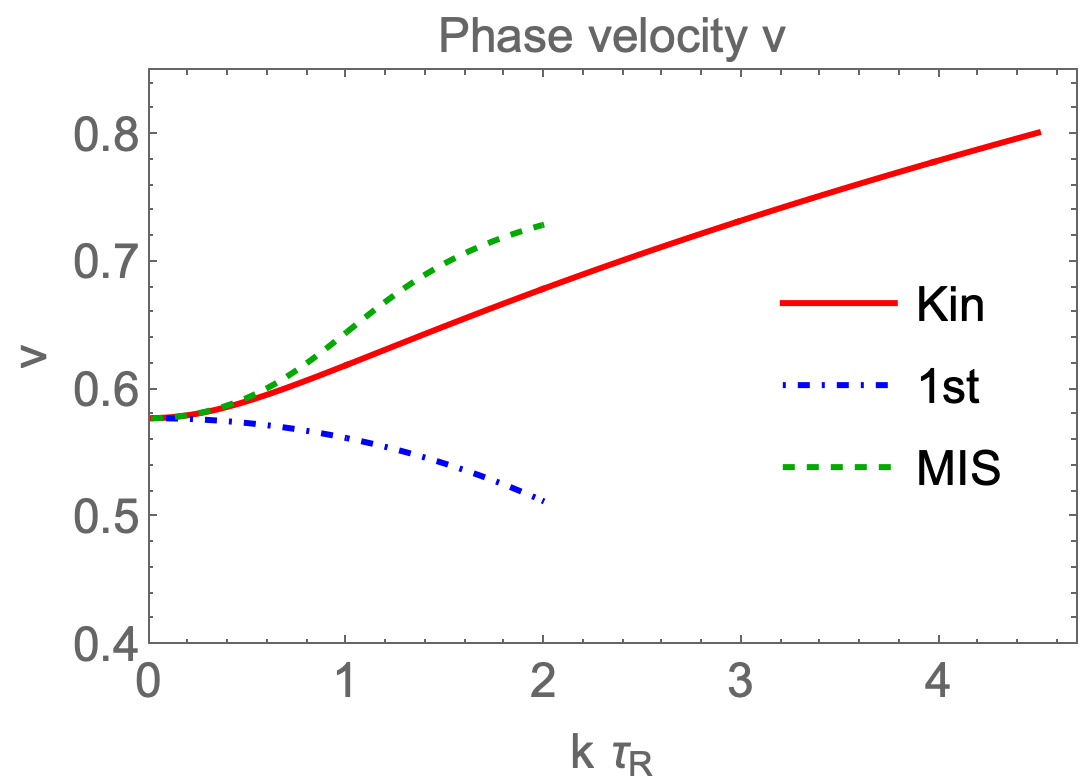}
    \includegraphics[width=.32\textwidth]{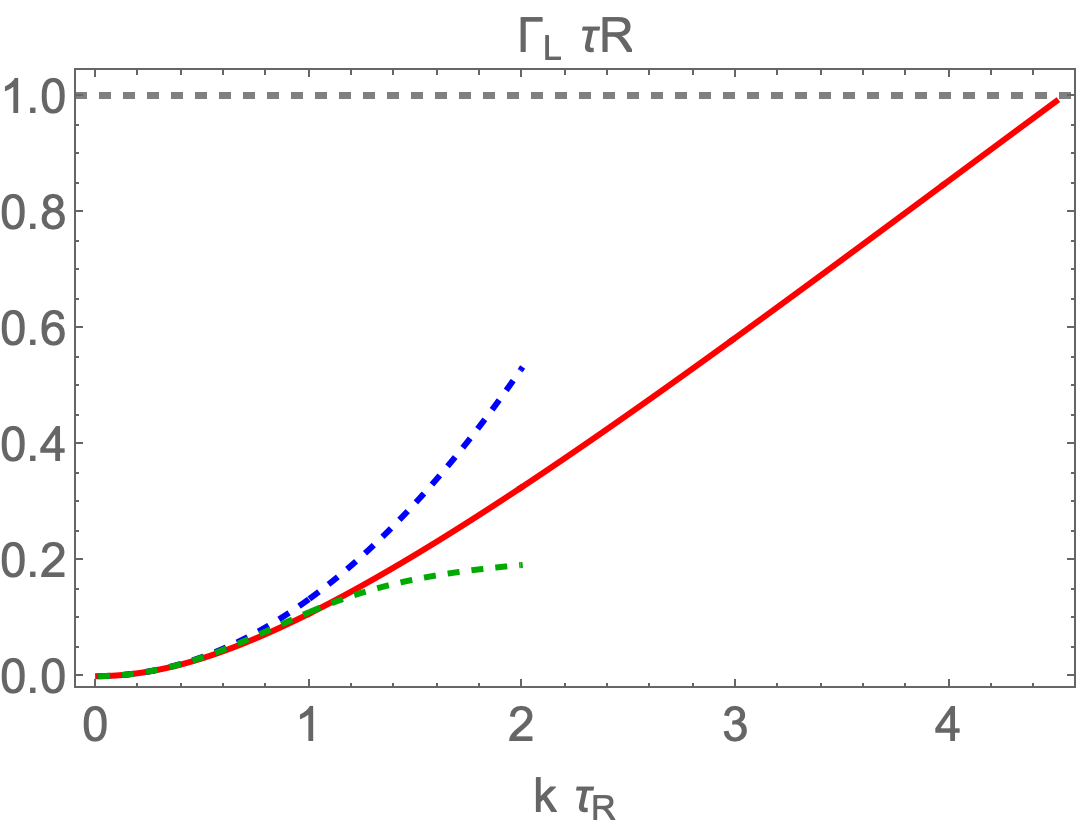}
        \includegraphics[width=.32\textwidth]{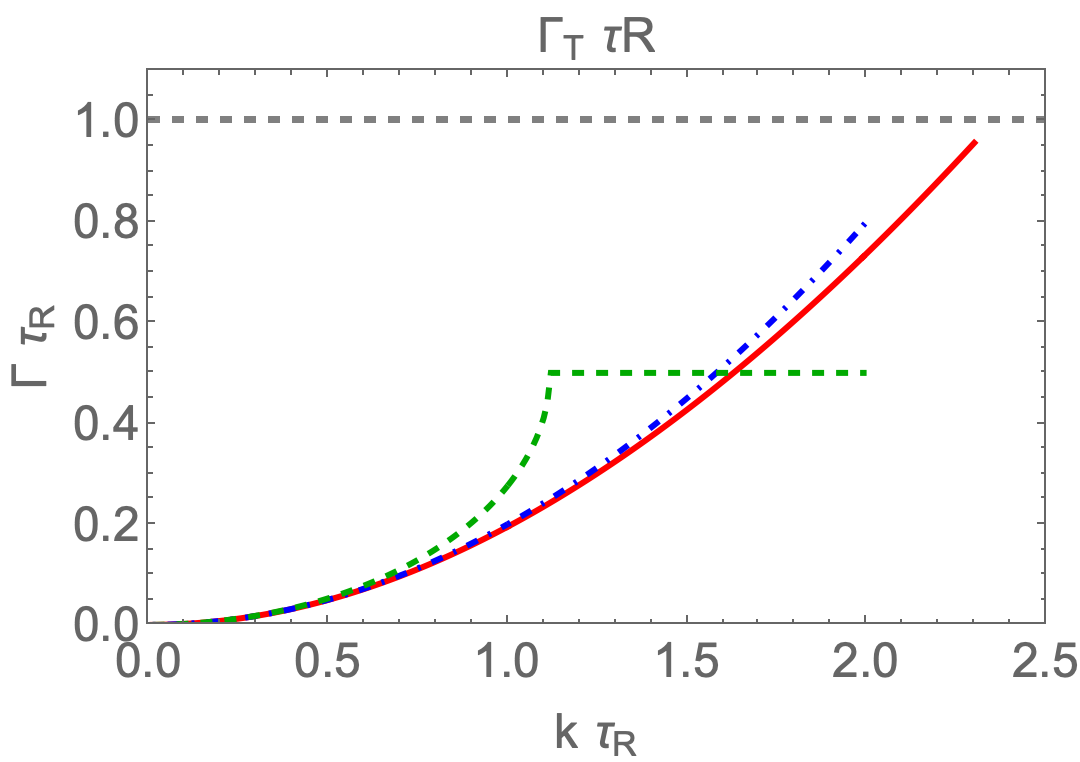}
  \caption{
        \label{fig:dispersion}
The dispersion of collective modes in RTA kinetic theory.
In the longitudinal channel, the solution to eq.~\eqref{sound-kin} represents a propagating mode $\o_{L}(k)=\pm v(k)\, k - i \Gamma(k)$. 
In the transverse channel, the solution to eq.~\eqref{shear-kin} is purely dissipative $\o_{T}(k)=-i \Gamma_{T}(k)$.
From left to right, the red curves in the Figure correspond to phase velocity $v$, attenuation rate $\Gamma$, and $\Gamma_{T}$ as a function of $\bark=k\tau_{R}$. 
For comparison, the first-order hydrodynamics and MIS theory results are plotted in blue dashed and green dotted curves, respectively. 
For Figures on the middle and left, the horizontal black dashed curves plot the damping rate of quasi-particle excitations, $1/\tau_{R}$. 
    }
\end{figure}

In Fig.~\ref{fig:RTA-decom}, which exhibits the RTA response function vs $\Delta t$ for three values of $\bark$ ($\bark<\bark_{O}$) in the longitudinal and transverse channel, we have shown the contribution from the poles and branch cuts separately. 
Remarkably, the collective modes have a prevalent impact on the behavior of response functions even at the non-hydrodynamic gradient, especially at sufficiently large $t$. 
To understand why this happens, we compared the damping rate of RTA collective modes with the relaxation rate of quasi-particle excitations shown in Fig.~\ref{fig:dispersion}. We refer to the difference between the two, $\Delta \G(k)=\tR^{-1}-{\rm Im}(\o(k))$, as the gap. If the gap is greater than zero, the contribution from quasi-particle excitations will be suppressed by $\exp(-\Delta \Gamma \Delta t)$, eventually making high-frequency hydrodynamic modes dominate the response. 
The gap remains open unless the value of $\bar{k}$ is close to $\bark_{O}$, as seen in Fig.~\ref{fig:dispersion}, which explains the importance of high-frequency sound and shear modes outside the hydrodynamic regime.

It is crucial to mention that the dispersion of those high-frequency collective modes can not be described by extrapolating the results in the usual hydrodynamic theories to a larger gradient. 
To elucidate this, we compute the poles of the hydrodynamic response functions~\eqref{GL-hydro},\eqref{GT-hydro} by solving 
\bes
\begin{align}
\label{sound-dis-MIS}
R(\o,k)=\o^{2}-c^{2}_{s}k^{2}+ i \nu(\o)\o k^{2}=0\, , 
\end{align}
in the longitudinal channel and
\begin{align}
\label{shear-dis-MIS}
\o+i \nu_{T}(\o)\ k^{2}=0\, . 
\end{align}
\ees
in the transverse channel. 
Here $\nu(\o)$ equals to $\nu_{0}$ for the first-order theory and is given by \eqref{nu-L-MIS} in MIS theory, and the relevant transport coefficients are listed in eq.~\eqref{H-para}. 
In Figure~\ref{fig:dispersion}, we confirm that first-order and MIS theory accurately describes the dispersion of the RTA collective modes in the hydrodynamic regime but generally fall short in capturing important qualitative features of the RTA modes as we move to a larger gradient, see also further discussion in Sec.~\ref{sec:static-com}.

\subsubsection{Extended hydrodynamic regime: generalities
\label{sec:EHR-gen}
}

More generally, one may introduce the notion of \textbf{"Extended Hydrodynamic regime"} (EHR), as discussed recently in Ref.~\cite{Ke:2022tqf}.
EHR has two defining properties: 1) high-frequency sound and/or shear modes exist that are gapped from other non-hydrodynamic excitations; 2) the dispersion relation of high-frequency sound and shear mode can not be described by conventional hydrodynamic theory. 
The RTA kinetic theory exemplifies the existence of EHR, as discussed in the preceding sections.

Importantly, EHR can appear in a wide range of microscopic theories.

First of all, 
the authors of Ref.~\cite{Du:2023bwi,Ochsenfeld:2023wxz} demonstrate remarkable agreement among the retarded correlator in the sound channel for kinetic theory with different collision kernels once time and gradient are rescaled by the effective relaxation time $\eta/w$. 
Those collision kernels include those obtained from RTA, scalar $\phi^{4}$, $SU(3)$ Yang-Mills theory and QCD effective kinetic theory with $N_{f}=3$.  
This study indicates that EHR, observed using the simplified RTA collision kernel, applies to more sophisticated and realistic collision kernels derived from quantum field theory.

Furthermore, high-frequency sound mode is not uncommon in strongly coupled gauge theories that can be calculated using holography duality. 
For example, in the strongly coupled ${\cal N}=4$ super-symmetric Yang-Mills theory in large $N_{c}$ limit (SYM), which captures many features of QCD in strong coupling, the sound mode exists at any finite gradient and has a smaller than damping rate than other excitations (quasi-normal modes)~\cite{Amado:2008ji,Fuini:2016qsc}. 
This similarity between kinetic theory and strongly coupled SYM theory is remarkable, given that the coupling strength is drastically different between the two.

Finally, various liquid metals can sustain sound modes extending from hydrodynamic regime to wavelengths comparable to inter-atomic distances, as reviewed in Refs.~\cite{RevModPhys.77.881,Trachenko_2015}. 
All the results discussed above indicate the generality of the EHR scenario.




\subsection{The construction of \MIS
\label{sec:MIS-STAR}
}

We review MIS* theory, which we proposed in Ref.~\cite{Ke:2022tqf}. 
The purpose of constructing MIS* is to provide a simple model that can describe the response in EHR.
Our approach is by "trial and error" strategy: we explore the simplest extension of MIS and evaluate its effectiveness. 
The resulting theory may be seen as a "deformation" of MIS theory that we decompose the non-equilibrium corrections to the EMT,$\pi^{\mu\nu}$~\eqref{pi-def}, into two parts:
\begin{align}
\label{pi-12}
    \pi^{\mu\nu}= \pi^{\mu\nu}_{1}+ \pi^{\mu\nu}_{2}\, , 
\end{align}
And we require the dynamic equation for $\pi_{1,2}$ to have the same form as eq.~\eqref{MIS} in MIS theory.:
\begin{align}
    \label{MIS-a}
    D\pi^{\mu\nu}_{a}=- \frac{1}{\tau_{a}}\le(\pi^{\mu\nu}_{a}+2\eta_{a}\, \sigma^{\mu\nu}\ri)\, , 
    \qquad 
    a=1,2\, ,
\end{align}
where $\eta_{1,2}, \tau_{1,2}$ are model parameters. 
Without losing generality, we shall assume that $\tau_{1}<\tau_{2}$, meaning $\pi_{2}$ 
relaxes faster than $\pi_{1}$.  
In the long time limit, $\pi^{\mu\nu}$ should approach to its value in the first-order hydrodynamics~\eqref{pi-vis},
yielding the following constraint
\begin{align}
\label{eta-sum}
    \sum^{2}_{a=1}\eta_{a}=\eta\, . 
\end{align}

We shall examine if MIS* can be used to describe EHR response in RTA kinetic theory. 
Following a similar procedure as in Sec.~\ref{sec:hydro-G}, we find $\sG_{L}, \sG_{T}$ from MIS* is of the same form as eqs.~\eqref{GL-hydro} and \eqref{GT-hydro}, respectively, with $\nu(\o), \nu_{T}(\o)$ given by
\begin{align}
\label{nu-KY}
    \nu(\o)= \frac{4}{3}\sum^{2}_{a=1}\frac{\nu_{a}}{1-i \o \tau_{a}}\, ,
    \qquad
    \nu_{T}(\o)= \frac{3}{4}\nu(\o)\, . 
\end{align}
where $\nu_{a}=\eta_{a}/w_{0}$.
To further simplify the model, 
we consider the limit $\tau_{2}/\tau_{1}\to 0$,
allowing eq.~\eqref{nu-KY} to be written as
\begin{align}
\label{nu-KY2}
    \nu(\o)= \frac{4\nu_{0}}{3}\le(\frac{1-\delta}{1-i\o\tau_{1}}+\delta\ri)\, . 
\end{align}
where two dimensionless ratios are defined for later convenience
\begin{align}
    \delta= \frac{\eta_{2}}{\eta}\, ,
    \qquad
    \gamma= \frac{\tau_{1}}{\tau_{\pi}}\, . 
\end{align}
This limit means that even at non-hydrodynamic timescales like $\omega\sim\tau^{-1}_{1}$, $\pi^{\mu\nu}_{2}$ has enough time to relax to $-2\eta_{2}\sigma^{\mu\nu}$ and contribute around $\propto \delta k^{2}$ to the damping rate of sound mode, which makes $\delta$ the parameter that controls the dissipative rate in EHR
\footnote{
Like MIS theory, MIS* is not causal for very large $k$ in the $\tau_{2}\to 0$ limit, but this is not a severe problem as long as we are working on the linear response regime. 
We can use a small but finite value of $r$ to ensure causality if necessary. 
}. 
We note that while $\tau_{\pi}$ in MIS theory is fixed by matching to the second-order hydrodynamics, we only require \MIS to reproduce the first-order hydrodynamic results and not higher-order corrections in the small gradient limit. 
With this little compromise, we could adjust $\tau_{1}$ or the ratio $\g$ to describe modes in EHR, making it an excellent trade-off, as we shall see shortly.

\subsection{Kinetic theory v.s. hydrodynamic theories}
\label{sec:static-com}

\begin{figure}
    \centering
    \includegraphics[width=0.32\textwidth]{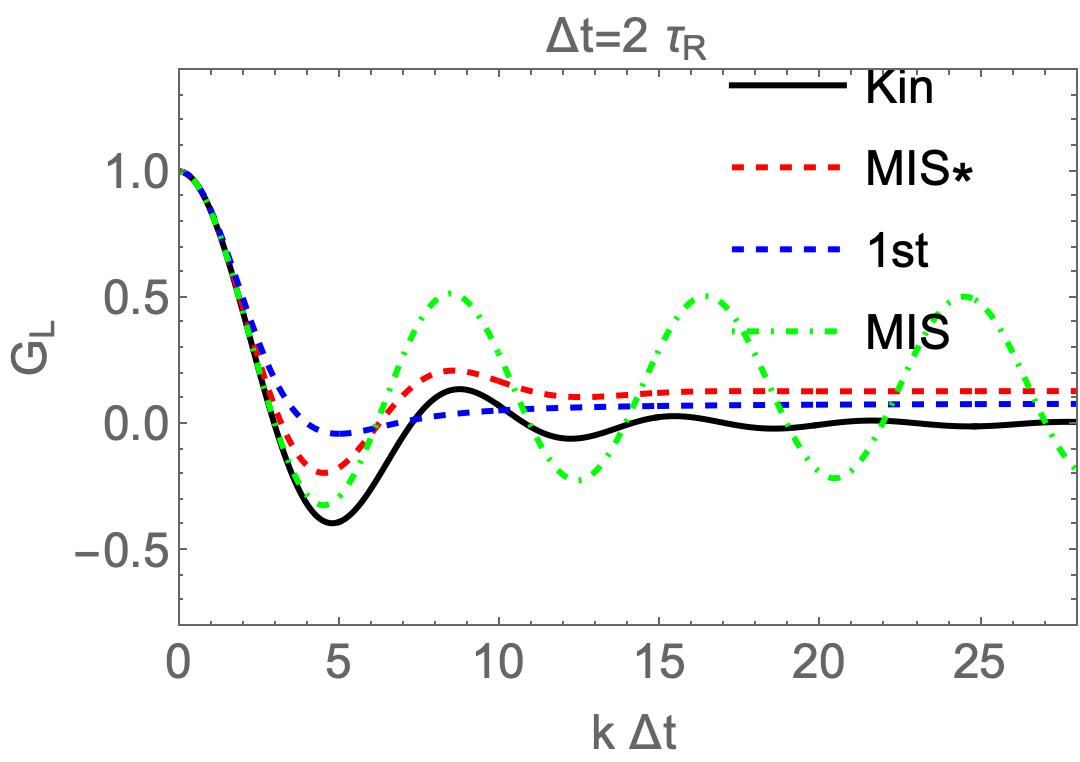}    
    \includegraphics[width=0.32\textwidth]{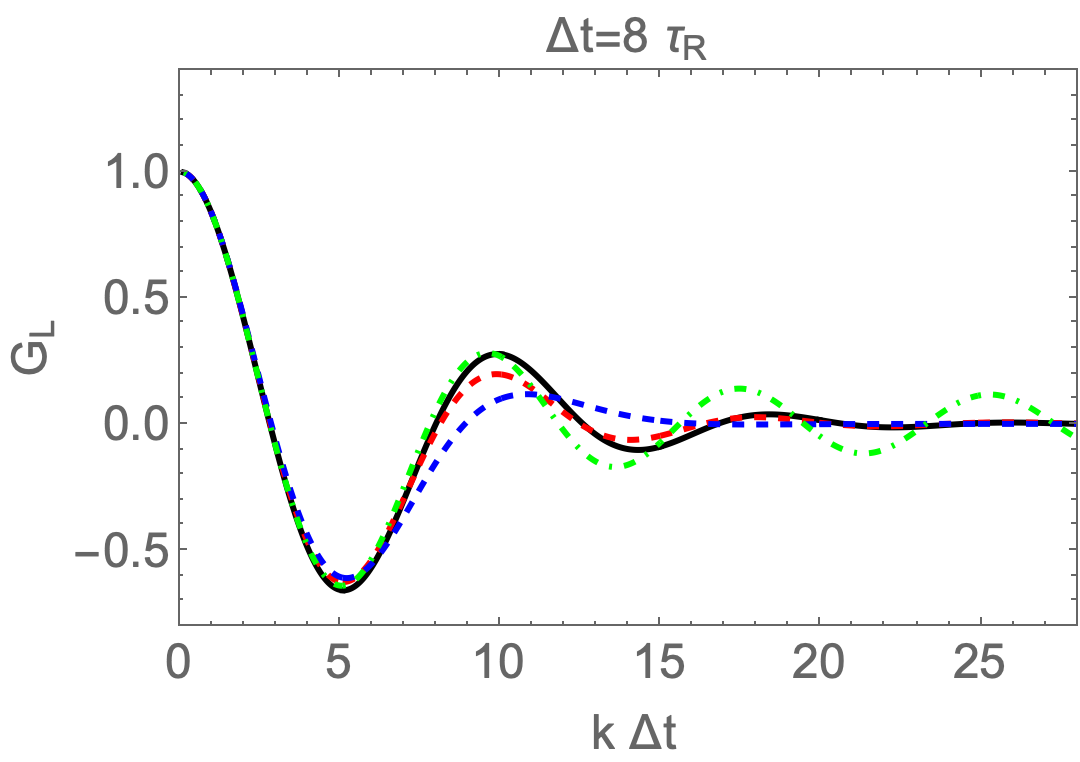}    
    \includegraphics[width=0.32\textwidth]{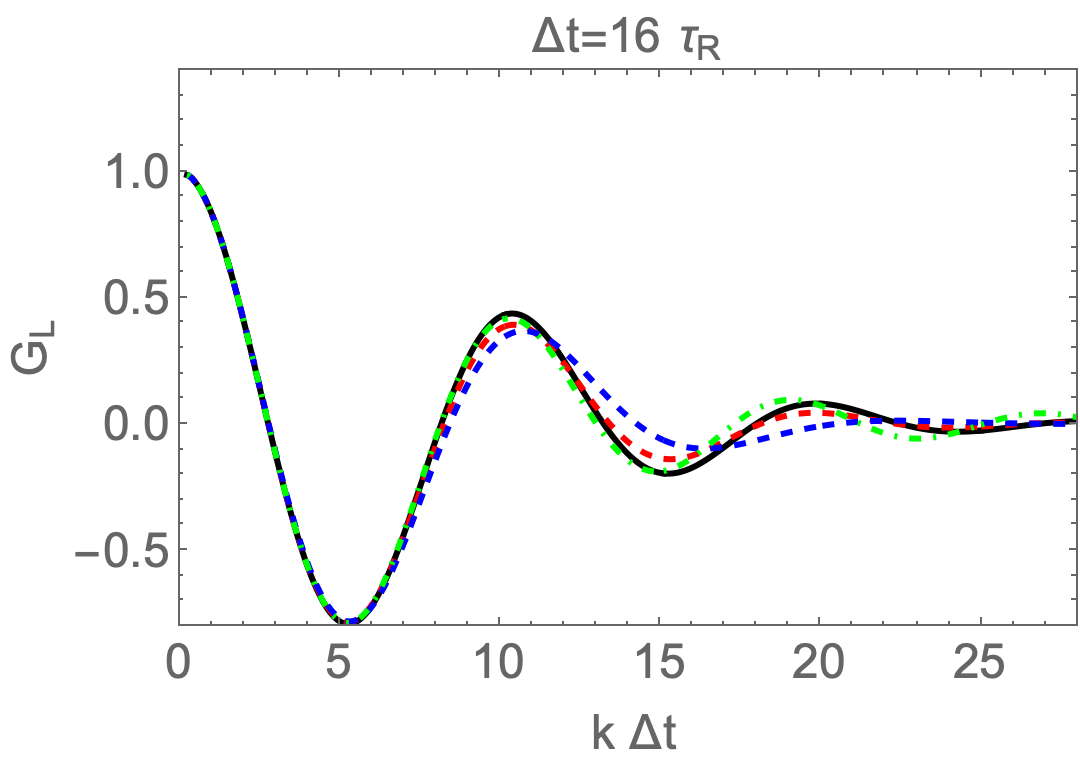}    
    \includegraphics[width=0.32\textwidth]{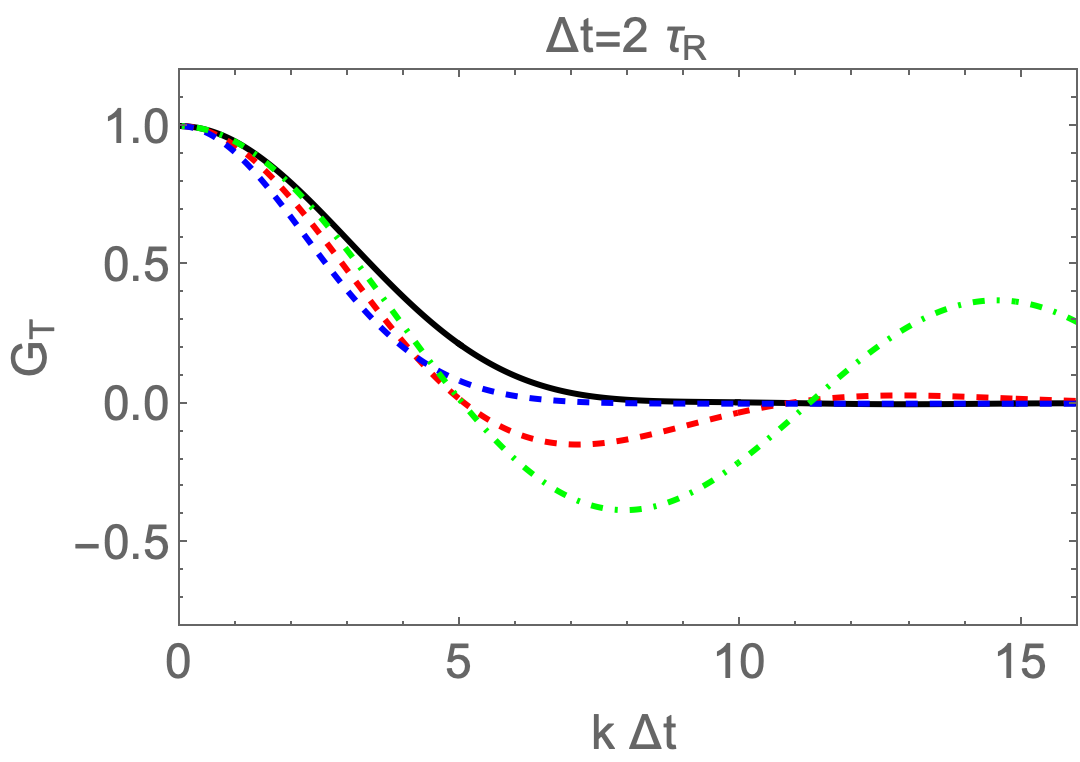}    
    \includegraphics[width=0.32\textwidth]{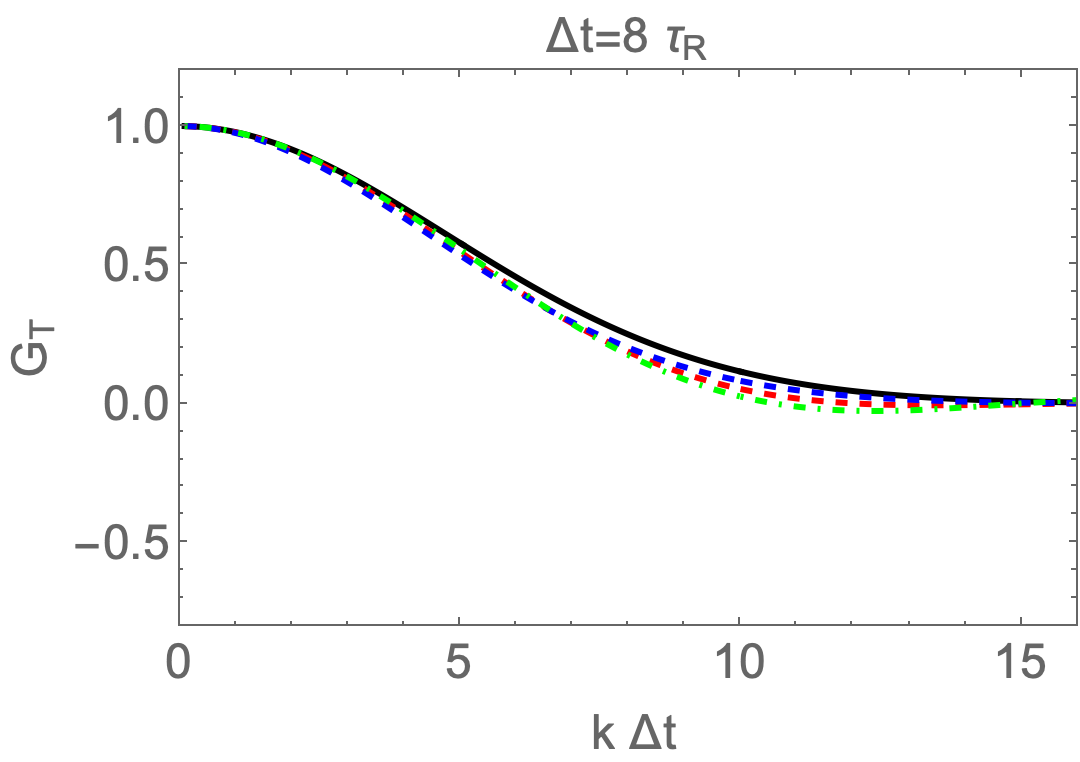}    
    \includegraphics[width=0.32\textwidth]{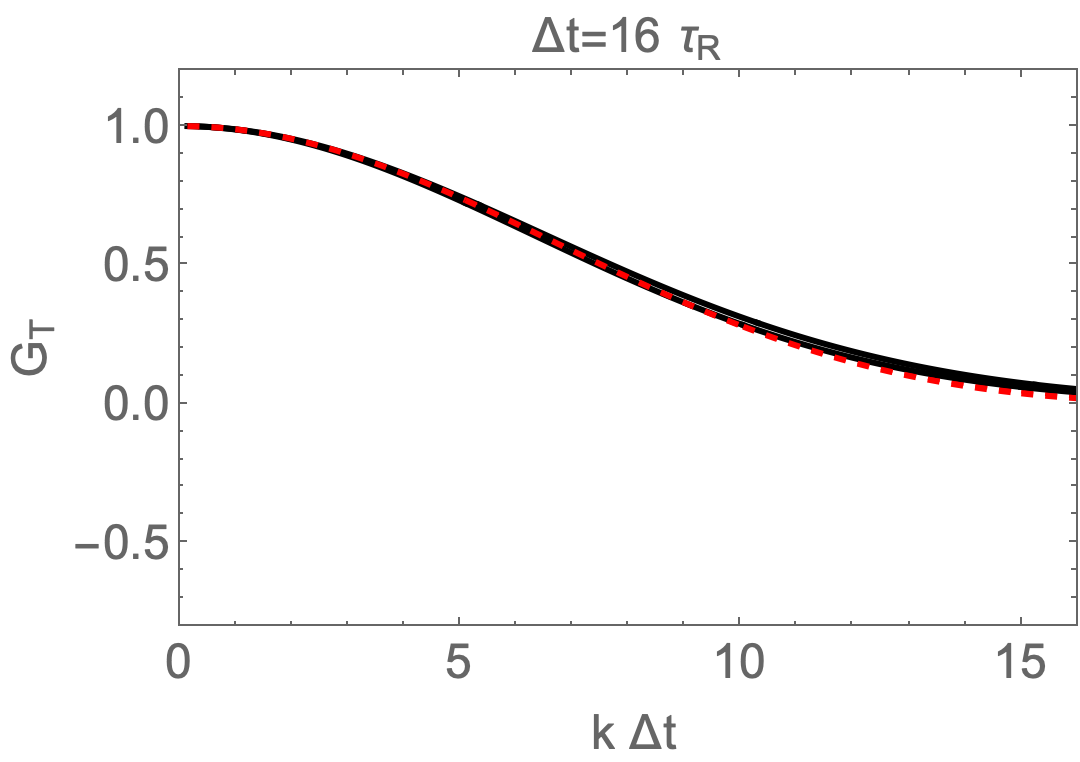}    
    \caption{
        \label{fig:RTA-response}
    The RTA response function vs $k\Delta t$ for the longitudinal channel (the upper panel) and the transverse channel (the lower panel).
    For comparison, MIS* (obtained by choosing $(\delta,\g)=(0.15,0.8)$), MIS, and the first-order hydrodynamic results are shown in red, blue, and green curves, respectively. 
    For their analogs in the Bjorken expanding background, see Fig.~\ref{fig:G_vs_k}. 
    }
\end{figure}


In this section, we will explore how the RTA response function behaves in a static and homogeneous equilibrium background in the time domain. 
To do so, we will use equation \eqref{G-t} alongside eqs.~\eqref{GL-kin} and \eqref{GT-kin}. 
We will then compare these results with their counterpart in the Bjorken expanding background in Sec.~\ref{sec:Bjorken-result}. 
Moreover, we will compare the RTA results with responses in "hydrodynamic theories," which include first-order MIS and MIS* theory. 
We put a quotation mark here to emphasize that some of those theories are to be extrapolated outside the hydrodynamic regime. 
The first-order and MIS theory results are obtained using parameters listed in eq.~\eqref{H-para}. 
For MIS*, we will use a suitable choice of $(\g, \delta)$, which could describe RTA sound dispersion; see Fig.~1 in Ref.~\cite{Ke:2022tqf}.

In Fig.~\ref{fig:RTA-response}, we show the longitudinal (the upper panel) and transverse (the lower panel) response function vs $k$ for $\Delta t=2\, \tau_{R}$ (left), 
$\Delta t=8\, \tau_{R}$ (middle) and  $\Delta t=16\, \tau_{R}$ (right). 
They represent early-time, intermediate-time, and late-time responses, respectively. 
For kinetic theory, $\sG_{L}$ exhibits the damped oscillation behavior while $\sG_{T}$ is purely dissipative. 
As $k$ increases, Hydrodynamic and RTA kinetic theory start to differentiate from each other. 
MIS theory, which includes higher-order gradient effects, only modestly improves the description when $k$ is small. 
Outside the hydrodynamic regime, both first-order and MIS theory are inadequate to describe the RTA response (see also Ref.~\cite{Hong:2010at}).
The discrepancy can be primarily explained by the difference in the dispersion of collective modes among those theories.
For instance, the phase velocity $v(k)$ of RTA sound mode rises from the conformal sound velocity $\sqrt{1/3}\approx 0.57$ up to $0.8$ as $k$ increases, and the sound attenuation rate also rises with $k$. 
In contrast, (extrapolated) first-order hydrodynamics does not give a supersonic phase velocity, and the sound damping rate grows faster than the RTA theory. 
Consequently, as shown in Fig.~\ref{fig:RTA-response},  first-order hydrodynamics is less oscillatory with a smaller amplitude than the RTA response function. 
As for the MIS theory, the phase velocity of the sound mode also increases, but its attenuation rate will be saturated instead of growing as $k$ becomes larger.
This explains the oscillating behavior in the MIS response function at large $k$ where RTA sound mode has already been damped out.

Suppose we define "hydrodynamization" of response as the point where the response function closely matches the hydrodynamic prediction at a specific $\Delta t$.\footnote{In many references,
"hydrodynamization" refers to the case when the bulk evolution is described by viscous hydrodynamics.
} 
In that case, we can see that in the shear channel, hydrodynamization occurs sooner than in the sound channel.
There are at least two reasons why this may occur.
The first reason is accidental:  the shear dispersion happens to be quantitatively similar in the first-order and RTA theory for a wide range of values for $k$. 
The second one is more general. 
The damping rate of the shear mode is greater than that in the sound channel at given $k$. 
Therefore, the non-hydrodynamic gradient will be damped first in the former sector. 
It is worthwhile mentioning that in the transverse channel, equation \eqref{shear-dis-MIS} has two roots at a fixed value of $k$. 
While both modes are imaginary for small $k$, they become a pair of propagating modes for $k$ larger than a certain threshold. 
This is responsible for the spurious wave-like behavior observed in the MIS response function in the shear channel.

In Fig.~\ref{fig:RTA-response}, we illustrate the comparison between RTA and MIS* response function with a representative combination of model parameters  $(\g, \delta)=(0.25,0.8)$.
Despite the model's simplicity, the agreement in the sound channel is rather remarkable. 
In the shear channel, the agreement is less impressive.  
Pursuing improvement in this channel is interesting, which we leave to future work.
That said, we stress that MIS* is still very useful in its current form when hydrodynamization happens first in the shear channel, since capturing the EHR response in the sound channel is more important in this case. 
This is what has been accomplished by MIS*.

\subsection{Discussion on the extension of hydrodynamics}

We developed the MIS* theory to describe how the medium responds in the extended hydrodynamic regime (EHR), where the high-frequency sound mode still has a longer lifetime than other excitations. 
Our analysis above shows that this framework could describe high-frequency hydrodynamic mode in EHR (or at least part of it) with a minimum number of model parameters. 
This is not the only possible formalism with which to describe EHR. 
An alternative method is to include higher-order gradient terms in hydrodynamics, and the efficiency of doing so relies on the convergence of the gradient expansion, which has attracted many recent studies~\cite{Grozdanov:2019kge,Grozdanov:2019uhi,Heller:2020hnq}. 
See also Ref.~\cite{Heller:2021yjh} for a different strategy in extending hydrodynamics.

In MIS*, $\pi_{1},\pi_{2}$ are introduced to describe sound dispersion outside the hydrodynamic regime. 
While including non-hydrodynamic fields to extend hydrodynamic is not new and has been widely discussed in literature~\cite{Heller:2014wfa,Gavassino:2023odx,Gavassino:2023qwl}, 
most of those studies are motivated by the expectation that the number of relevant modes increases at shorter time scales or non-hydrodynamic distances. 
Moreover, the emergence of parametrically slow modes also necessitates including additional modes even in the hydrodynamic regime, and the result theory is referred to as "Hydro+" in Ref.~\cite{Stephanov:2017ghc}. 
One familiar situation is near a critical point; the relaxation time of the fluctuations of the order parameter field grows with increasing correlation length. 
In all cases discussed above, the number of relevant modes is larger than that of hydrodynamic ones, which are conserved densities.  
In contrast, the number of slow modes in EHR is the same as that of hydrodynamic ones in the sound channel.
The purpose of including non-hydrodynamic dynamic fields $\pi^{\mu\nu}_{1,2}$ is not because they are additionally slow modes but because they bring about new parameters needed to characterize the dispersion of high-frequency sound modes. 
\footnote{
In our view, to understand the need to introduce non-hydrodynamic d.o.fs. $\pi^{\mu\nu}_{1,2}$ in MIS*, it might be useful to recall that some heavy particles are introduced to make neutrino light through the sea-saw mechanism in some beyond-standard model scenarios.
}
This distinguishes the \MIS theory from early studies.

\section{Perturbation in medium undergoing Bjorken expansion
\label{sec:B-formalism}
}

In this section, we will study the energy-momentum response around a Bjorken expanding conformal plasma and compare the results with those in the static background.

\subsection{Response function for Bjorken background
\label{sec:response-B}
}

We divide the stress-energy tensor (EMT) into a background and a perturbation part:
\begin{align}
    T^{\mu\nu}(\tau,\vx)=T^{\mu\nu}_{0}(\tau)+\delta T^{\mu\nu}(\tau,\vx)\, ,
\end{align}
where we assume the background is under Bjorken expanding. 
Accordingly, background EMT $T^{\mu\nu}_{0}(\tau)$ only depends on the Bjorken proper time $\tau=\sqrt{t^{2}-z^{2}}$ and takes form 
\begin{align}
\label{EMT-bg}
T^{\mu\nu}_{0}={\rm diag}\le(\e_{0}, P_{0}, P_{0},P_{0}\ri)
+{\rm diag}\le(0,1 , 1,-\frac{2}{\tau^{2}}\, \ri)\,\pi_{0} \, .
\, 
\end{align}
Here, the first and second terms on the R.H.S. of eq.~\eqref{EMT-bg} represent the contribution from the ideal and non-equilibrium part of the constitutive relation, see eq.~\eqref{pi-def}. 
The background $\pi^{\mu\nu}_{0}$ is diagonalized and is fully parameterized by $\pi_{0}=\pi^{xx}_{0}=\pi^{yy}_{0}$ for a conformal system under consideration.

Following Ref.~\cite{Kurkela:2018vqr}, we define the response function describing the subsequent evolution of the initial perturbation $\delta T^{\mu\nu}$ at $(\tau', \vx'_{\perp})$ (c.f. eq.~\eqref{Response-def}):
\begin{align}
    \delta T^{\mu\nu}(\tau,\vx_{\perp})= \int d^{2}\vx'_{\perp}\, \sG^{\mu\nu}_{\,\,\,\,\,\a\b}(\vx_{\perp}-\vx'_{\perp};\tau,\tau')\, \d T^{\a\b}(\vx'_{\perp},\tau')\, .
\end{align}
where we have used the fact that the response functions depend only on the different $\vr=\vx_{\perp}-\vx'_{\perp}$. 
For the present purpose, 
we only consider the perturbation that depends on the spatial vector lying entirely in the plane transverse to $z$-direction $\vx_{\perp}$.
The non-zero independent components of $\delta T^{\mu\nu}$ are 
\begin{align}
\label{non-zero-dEMT}
\delta \e=\delta T^{\tau\tau}\, 
\qquad
\delta g^{a}\equiv\delta  T^{\tau a}\, , 
\qquad
\delta T^{ab}\, , 
\end{align}
where $a=x,y$ and $\delta g^{a}$ is nothing but transverse momentum density perturbation. 
The value of $\delta T^{\eta\eta}$ is related to $\delta \e, \delta T^{ab}$ by the conformality condition. 
We are interested in the response of the plasma near equilibrium, so we shall implicitly assume that $\tau'$ is larger than the thermalization time of the plasma. 
The response functions of a far-from-equilibrium Bjorken expanding plasma have been studied using effective QCD kinetic theory~\cite{Kurkela:2018vqr,Kurkela:2018wud,Du:2020zqg} and RTA kinetic theory~\cite{Kamata:2020mka}.

Following the discussion of Ref.~\cite{Kurkela:2018vqr}, one can show that there are five independent components of $\sG^{\mu\nu}_{\,\,\,\,\,\a\b}$ that describe the energy-momentum response to the initial energy density $\delta \e$ and initial transverse momentum density  $\delta g^{a}$ disturbance (see also~Ref.~\cite{Kamata:2020mka}):
\bes
\label{G-list}
\begin{align}
&\,    \sG^{\tau\tau}_{\,\,\,\,\,\tau\tau}(\vr)\equiv {\cal G}_{\epsilon\epsilon}(r)\, , 
\\
&\,
\sG^{\tau a}_{\,\,\,\,\,\tau\tau}(\vr)\equiv \hat{r}^{a}\, {\cal G}_{\e L}(r)
    \qquad
  \sG^{\tau\tau}_{\,\,\,\,\,\tau a}(\vr)\equiv\hat{r}^{a}\, {\cal G}_{L\e}(r)\, ,
    \\
&\,  \sG^{\tau a}_{\,\,\,\,\,\tau b}(\vr)= \hat{r}^{a}\hat{r}^{b}{\cal G}_{LL}(r)+\le(\delta^{ab}- \hat{r}^{a}\hat{r}^{b}\ri){\cal G}_{TT}(r)\, . 
\end{align}
\ees
In the above expressions, we suppress the dependence on $\tau,\tau'$ in \RF to highlight their dependence on $\vr$. 
The physical interpretation of those \RF should be transparent. 
For example, ${\cal G}_{\epsilon L}$ describes energy density at $\tau$ induced by the momentum density disturbance projected along direction $\hat{r}$ at initial time $\tau'$. 
Likewise, 
${\cal G}_{TT}$ describes the transverse (with respect to $\hat{r}$) momentum density induced by the initial momentum density.

Following the common practice (see Refs~\cite{Kurkela:2018wud,Kamata:2020mka,Du:2020zqg}), we first compute $\sG^{\mu\nu}_{\,\,\,\,\,\a\b}$ in Fourier space $\vk=(k_{x},k_{y})$ and obtain the real space \RF 
\begin{align}
\label{RF-fourier}
    \sG^{\mu\nu}_{\,\,\,\,\,\a\b}(\vx-\vx_{0};\tau,\tau_{0})= \int \frac{d^{2}\vk}{(2\pi)^{2}}\,\sG^{\mu\nu}_{\,\,\,\,\,\a\b}(\vk;\tau,\tau_{0}) \, e^{i\vk\cdot (\vx-\vx_{0})}\, . 
\end{align}
The relevant independent components of $G^{\mu\nu}_{\,\,\,\,\,\a\b}(\vk;\tau,\tau_{0})$ are~\cite{Kurkela:2018vqr}
 \begin{eqnarray}
 \label{Gk-def-B}
\sG^{\tau\tau}_{\,\,\,\,\,\tau\tau}(\vk) &=&
    {\cal G}_{\e\e}(\vk)\, , \\
\sG^{\tau \tau}_{\,\,\,\,\,\tau a }(k)&=& i\, \hk_{a}  {\cal G}_{\e  L}(k)\, , 
  \qquad
\sG^{\tau a}_{\,\,\,\,\,\tau\tau}(\vk) = i\, \hk_{a}\, {\cal G}_{L \e}(k),
 \\
\sG^{\tau a}_{\,\,\,\,\,\tau b}(\vk)&=& \hk^{a}\hk^{b}\,{\cal G}_{LL}(k)+(\delta^{ab}-\hk^{a}\hk^{b})\, {\cal G}_{TT}(k).
 \end{eqnarray}
They determine the real-space response function through the relation:
\bes
\label{eq:response:Gk-Gr}
\begin{eqnarray}
\sG_{\e \e}(r) &=& \int \frac{dk k }{2\pi}\,J_0(kr)\,\sG_{\e \e}(k)\, , \\
 \sG_{ \e L} (r) &=& \int \frac{dk k}{2\pi}\,J_1(kr) \sG_{\e L}(k)\, ,
 \qquad
 \sG_{L\e}(r) = \int \frac{dk k}{2\pi}\,J_1(kr) \sG_{L\e}(k)\, ,
 \\
\sG_{LL}(r) &=& \int \frac{dk k }{2\pi} \frac{1}{2}\left[\le(J_0(kr)-J_2(kr)\ri)\, \sG_{LL}(k) + \le(J_0(kr)+J_2(kr)\ri) \sG_{TT}(k)\right]\, , 
 \\
\sG_{TT}(r) &=& \int \frac{dk k }{2\pi} \frac{1}{2}\,\left[\le(J_0(kr)+J_2(kr)\ri) \sG_{LL}(k) + 
\le(J_0(kr)-J_2(kr)\ri) \sG_{TT}(k)\right]\, ,
\end{eqnarray}
\ees
where $J_{0,1,2}$ are standard Bessel functions.

\subsection{Kinetic theory calculation
\label{sec:kin-B}
}

In this section, we present the calculation of the response function for a Bjorken expanding plasma using RTA kinetic equation~\eqref{RTA}. 
In Bjorken coordinate 
the kinetic equation~\eqref{RTA} reads
\begin{eqnarray}
\label{eq:rta}
\frac{\partial f}{\partial \tau} + \frac{\mathbf{p}_{a}}{p}\frac{\partial f}{\partial x_{a}} - \frac{p^\eta}{\tau} \frac{\partial f}{\partial p^\eta}= 
-\frac{p_{\mu} u^{\nu}\,(f-f_{\rm eq})}{p\,\tau_{R}}
\end{eqnarray} 
where the summation over the dummy variable $a$ is understood. 
As explained earlier, we consider massless particles with single-particle energy $p=|\vp|$.

Since we are interested in energy-momentum disturbance, it is convenient to introduce the integrated distribution function (see also Refs.~\cite{Brewer:2019oha,Brewer:2022ifw})
\begin{align}
\label{F-def}
F(\hp) = \int^{\infty}_{0} \frac{dp}{2\pi^{2}}\,  p^{3} f\, ,
\end{align}
from which one can obtain the components of EMT through the solid angle integration with the appropriate weight, e.g.,
\begin{align}
\label{F-EMT}
\e =\int \frac{d\Omega}{4\pi}\, F, 
\qquad 
g_{a}=\int \frac{d\Omega}{4\pi} \frac{p_{a}}{p}\,F\, ,
\qquad
T_{ab}=\int \frac{d\Omega}{4\pi} (\frac{p_{a}p_{b}}{p})\, F\, . 
\end{align}
By performing the integration over $dp$ over the kinetic equation \eqref{eq:rta}, we then derive the equation for $F$ (see also Refs.~\cite{Brewer:2019oha,Kurkela:2019kip,Brewer:2022vkq}) 
\begin{align}
\label{F-evo}
\le[\pd_{\tau}  + \hat{p}_{a}\frac{\pd}{\pd x_{a}}+
\frac{1}{\tau}
\le(
4\cos^2\theta
-\cos\theta (1-\cos^2\theta
\ri)
\pd_{\cos\theta}
\ri]\, F
= - \frac{1}{\tau_{R}} \int^{\infty}_{0} \frac{dp}{2\pi^{2}}\,  p^{2}\,(p\cdot u)\, (f-f_{eq})\, . 
\end{align}

We first consider background solution $F_{0}(\tau,\cos\theta)$ with $u^{\mu}_{0}=(1,0,0,0)$(in Bjorken coordinate). 
The equation~\eqref{F-evo} reduces to 
\begin{align}
\label{F0}
\left[ \partial_\tau +
\frac{1}{\tau}
\le(
4\cos^2\theta
-\cos\theta (1-\cos^2\theta
\ri)
\pd_{\cos\theta}
 \right] F_{0}(\tau,\cos\theta) &= \frac{1}{\tau_R} \le[\e_{0}(\tau) - F_{0}(\tau,\cos\theta)\ri]\, .  
\end{align}
The R.H.S. of eq.~\eqref{F0} can be understood from the fact that in equilibrium, $F_{0}=\e_0$ and is independent of $\cos\theta$.

Next, we turn to the perturbation around the background solution $F=F_{0}+\delta F$.
Linearizing the L.H.S. of eq.~\eqref{F-evo} is straightforward. 
For the R.H.S., we have
\begin{eqnarray}
\label{delta-C}
\frac{1}{p}\,\delta\le[
\frac{p\cdot u}{\tau_{R}}\, (f-f_{\rm eq})
\ri]
 &= \frac{\delta f-\delta f_{\textrm{eq}}}{\tau_{R,0}}
 +
 \left(p\cdot \delta u - \frac{p^{0}\, \delta \tau_{R}}{\tau_R}\right)\,  \frac{f_{0}-f_{\textrm{eq},0}}{\tau_{R,0}}\, .
\end{eqnarray}
Note $\delta \tau_{R}$ term will be present because $\tau_{R}$ depends on $\e$ (see eq.~\eqref{tau-scale}). 
For convenience, we shall count $f_{0}-f_{\rm eq,0}$ of the order ${\cal O}(\delta_{0})$ where $\delta_{0}$ is assumed to be another small parameter as we are interested in the near equilibrium background. 
In fact, $\pi_{0}/w_{0}\sim {\cal O}(\d_{0})$ in this counting scheme.
For a simple estimate of this ratio, we use the first-order hydrodynamic expectation $\pi_{0}/w_{0}= 2\eta /(3\tau w)$. 
Using $\eta/w=(1/5) \tau_{R}$ in the RTA kinetic theory~\eqref{H-para}, 
we find $\pi_{0}/w_{0}\leq 2/15$ for $\tau>\tau_{R}$ and is numerically small. 
See also Fig.~\ref{fig:BG} where the numerical solutions to the background equation~\eqref{F0} is plotted and the smallness of $\pi_{0}/w_{0}$ is verified. 
To further simplify our analysis, we will ignore the terms of the order ${\cal O}(\d_{0}\,\d)$, i.e. the second term on the R.H.S of eq.~\eqref{delta-C}.

To calculate the first term on the R.H.S of eq.~\eqref{delta-C}, one can use an expression analogous to eq.~\eqref{delta-f-eq} to expand $\delta f_{{\rm eq}}$. 
On the other hand,
\begin{align}
\delta T^{\tau a}=\delta g^{a}= w_{0}\delta u^{a}+\delta \pi^{\tau a}\, , 
\end{align}
and  $\pi^{\mu\nu}$ is transverse to the fluid velocity
\begin{align}
 \delta u_{\mu} \pi^{\mu\nu}_{0} +  u_{\mu}\delta \pi^{\mu\nu}_{0} = 0\, , 
\end{align}
so we find $\delta \pi^{\tau a} = \pi_0\delta u^a$ and consequently
\begin{align}
\label{du-g}
\delta u^{a}= \frac{g^{a}}{w_{0}(1-\frac{\pi_{0}}{w_{0}})}=
\frac{g^{a}}{w_{0}}+{\cal O}(\delta\,\delta_{0})\, . 
\end{align}
After integrating the linearized kinetic equation over $\int^{\infty}_{0}dp\, p^{2}$, 
we obtain, in Fourier space, the equation for $\delta F$ (see also Ref.~\cite{Brewer:2022vkq})
\begin{align}
\label{lin-F}
\le[ \partial_\tau + i\,k_{a}\,p_{a}+
\frac{1}{\tau}
\le(
4\cos^2\theta
-\cos\theta (1-\cos^2\theta
\ri)
\pd_{\cos\theta}
\ri]\delta F  &= \frac{1}{\tau_R}
\left( \delta \e+\frac{1}{c^{2}_{s}}\delta g^{a}\cdot \hat{p}
- \delta F\right)\, .
\end{align}
By solving eq.~\eqref{lin-F}, we can obtain the response function through the definition in Sec.~\ref{sec:response-B}. 

\subsection{Hydrodynamic theory calculation
\label{sec:hydro-B}
}

As we did in the previous sections~\ref{sec:hydro-G},
we collectively call first-order hydrodynamics, MIS (or its variants), and MIS* "hydrodynamic theories."

We shall first derive E.o.M for background energy density $\e_{0}$ and linearized energy and momentum density $\delta \e, \delta g^{a}$. 
The resulting equations apply equally to all "Hydrodynamic theories" under consideration. 
In contrast, the description of $\pi^{\mu\nu}$ differs among those theories and will be specified later.

We begin with the energy-momentum conservation equation
\begin{align}
\label{EM-con-B}
    \nabla_{\mu}T^{\mu\nu}= \pd_{\mu}T^{\mu\nu}+ \Gamma^{\mu}_{\mu\lambda}T^{\lambda\nu}+\Gamma^{\nu}_{\mu\lambda}T^{\mu\lambda}\, , 
\end{align}
where the non-zero connection in Bjorken coordinates are $\Gamma^{\eta}_{\eta\eta}=\tau, \Gamma^{\eta}_{\eta\tau}=\Gamma^{\eta}_{\tau\eta}=1/\tau$.
Using the expression for the background EMT~\eqref{EMT-bg}, we obtain
\begin{align}
    \pd_{\tau}\e_0 = - \frac{4}{3\tau} \e_0 + \frac{2\pi_0}{\tau}\ , 
\end{align}
Turning to the perturbation, 
we have from~\eqref{EM-con-B} that
\bes
\label{lin-B}
\begin{align}
  \left(\pd_{\tau}+\frac{4}{3\tau}\right)\delta \e + \partial_a\delta g^a  
  - \tau\,\delta\pi^{\eta\eta} = 0,\\
  \left(\pd_{\tau}+\frac{1}{\tau}\right)\delta g^a + \frac{1}{3}\partial^a \delta \e + \partial_b\delta \pi^{ba} = 0, \, ,
\end{align}
\ees
where we have used eq.~\eqref{non-zero-dEMT}.

Now, we turn to discuss $\pi^{\mu\nu}_{0}, \delta \pi^{\mu\nu}$ for different "hydrodynamic theories". 
As discussed earlier,
the background $\pi^{\mu\nu}_{0}$ is fully determined by $\pi_{0}$.
To close eqs.~\eqref{lin-B}, 
only $\delta \pi^{ab}$ are needed since $\d \pi^{\eta\eta}$ can be expressed in terms of $d\pi^{ab}$ due to the conformality condition, i.e.,
\begin{align}
    \delta \pi^{\eta\eta}=-\frac{1}{\tau^{2}}\sum_{a}\delta \pi^{aa}\, . 
\end{align}

In the first-order hydrodynamics, $\pi^{\mu\nu}$ is not dynamical and \eqref{pi-vis} yields
\begin{align}
\pi_0=
-2\eta_{0}\, \sigma^{\mu\nu}_{0}\, , 
\qquad
\delta \pi^{\mu\nu}= -2\eta_{0} \delta\sigma^{\mu\nu}\, .
\end{align}
where for the background
\begin{align}
    \sigma^{\mu\nu}_{0}=\frac{1}{3\tau}\textrm{diag}\{0, -1,-1,\frac{2}{\tau^2}\}\, . 
\end{align}
Further using eq.~\eqref{du-g}, we arrived at the expression:
\begin{align}
   \delta\sigma^{ab} =\frac{1}{w_{0}}\, \pd^{\langle a}\,\delta g^{b\rangle} \, . 
\end{align}

Following Ref.~\cite{Jaiswal:2019cju}, 
we shall use a modern variant of MIS, i.e. the theory formulated by Denicol, Niemi, Molnar and Rischke (DNMR)~\cite{Denicol:2012cn} where the equations of motion for $\pi^{\mu\nu}$ reads (see Ref.~\cite{Baier:2007ix} for examples of other variants of the second-order hydrodynamic theories ):
\begin{align}
\label{DNMR}
    D \pi^{\mu\nu}= - \frac{\pi^{\mu\nu}+2\eta \s^{\mu\nu}}{\tau_{\pi}} - \frac{4}{3}\pi^{\mu\nu}\theta-\tilde{\lambda}_1\pi_\a^{\langle\mu}\sigma^{\nu\rangle\a} - \ldots \, . 
\end{align}
Here, $\ldots$ on the L.H.S of \eqref{DNMR} denotes a term proportional to the vorticity, but we shall discard this contribution since the Bjorken flow is irrotational. 
The second-order transport coefficient equals $\tilde{\lambda}_1 = 10/7$ by matching to RTA kinetic theory~\cite{Jaiswal:2019cju}.

The DNMR equation for the background reads~\cite{Jaiswal:2019cju}
\begin{align}
\label{DNMR-B}
    \pd_{\tau}\pi_{0}= - \frac{1}{\tau_{\pi}}\,
    \left(\pi_{0}-\frac{2\eta}{3\tau}\right)- \left(\frac{4}{3}+\frac{\tilde{\lambda}_{1}}{3}\right)\,\frac{\pi_{0}}{\tau}
\end{align}
and the linearized equation can be written as
\begin{align}
\label{DNMR-lin}
\pd_\tau\delta\pi^{ab}  
 =&- \frac{\delta\pi^{ab}+2\eta\delta\sigma^{ab}}{\tau_\pi} + \frac{4}{3\tau} \delta\pi^{ab} +{\cal O}(\delta_{0}\delta)\, . 
 \end{align}
Here and hereafter, we shall drop the terms of the order ${\cal O}(\delta_{0}\delta)$ as we have already done in kinetic theory calculations.

Finally, 
we discuss the MIS* theory introduced in Sec.~\ref{sec:MIS-STAR} in the limit $\tau_{2}/\tau_{1}\to 0$ such that 
$\pi^{\mu\mu}_{2}$ (see eq.~\eqref{pi-12})
is given by
\begin{align}
\pi^{\mu\nu}_{2} = -2\eta_{2}\sigma^{\mu\nu}\, .     
\end{align}
For $\pi^{\mu\nu}_{1}$, instead of using \eqref{MIS-a}, we consider the equation of the form~\eqref{DNMR}, i.e.,
\begin{eqnarray}
\label{MIS-2}
\pi^{\mu\nu} &=& \pi_1^{\mu\nu}+\pi_2^{\mu\nu},\\
D \pi^{\mu\nu}_{1} &=& -\frac{\pi^{\mu\nu}+2\eta_{1} \s^{\mu\nu}}{\tau_{1}} -\frac{4}{3}\pi^{\mu\nu}_{1}\theta- \tilde{\l}_1\pi^{\langle\mu}_{\,\,\,\a}\s^{\nu\rangle\a}\,  
\end{eqnarray}
so that when $\eta_{2}=0$, our theory reduces to eq.~\eqref{DNMR}. 
Consequently, the background and linearized equation for $\pi^{\mu\nu}_{1}$ can be obtained by replacing $\pi_{0}, \delta \pi^{ab},\eta$ etc  in eqs.~\eqref{DNMR-B}, \eqref{DNMR-lin}. with $(\pi_{1})_{0}, \delta \pi^{ab}_{1}, \eta_{1}$ respectively.

\section{Results for Bjorken expansion
\label{sec:Bjorken-result}
}

\subsection{Initial condition}

We begin the background evolution at $\tau \geq \tau_R(\e(\tau))$ when the system is expected to be near equilibrium. 
Because of the conformality of the system under consideration, 
the solutions to eq.~\eqref{F0} and \eqref{lin-F} only depend on the rescaled time $\tau/\tau_{R}$. 
Instead of specifying the overall normalization in $\tau_{R}\propto \e^{-1/4}$, 
we first define $\tau_{0}$ such that $\tau_{0}=\tau_{R}(\e_{\rm in})$ for a given $\e_{\rm in}$. 
We then use $\tau_{0}$ to set up the temporal scale and present results in terms of $\tau/\tau_{0}$ or $\tau/\tau_{R}$. 
By construction, the background evolution starts at $\tau/\tau_{0}=1$

Since the background evolution of "hydrodynamic theories" is fully determined by the initial values of energy density $\e_{\in}$ and that of $\pi_{0}$, denoted by $\pi_{\in}$, 
we require the initial profile of $F$ is also fully specified by $\e_{\in}, $$\pi_{\in}$ to facilitate a comparison. 
According to eq.~\eqref{F-EMT}, we use the parameterization:
\begin{align}
F_{\in}(\cos\theta) = \e_{\in}+ 
\frac{15}{4}(1-\cos^{2}\theta)\,\pi_{\in}\, . 
\end{align}
We use the same $\e_{\in},\pi_{\in}$ for the first order and MIS theory. 
For MIS*, the dynamical field is $\pi^{\mu\nu}_{1}$ and will be initialized as $\pi^{\mu\nu}_{1}(\tau_{0})=(1-\delta)\pi_{\in}$.

After obtaining the background evolution, we then turn on initial energy disturbance  $\delta\e_{\in}$ (or momentum density $g^{x}_{\in},g^{y}_{\in}$) at $\tau'=2 \tau_{0}$. 
Correspondingly, we set $\d F(\theta)= \delta\e_{\in}$ (or $\d F(\theta)= \delta g^{x}_{\in}\sin\theta\cos\phi,\delta g^{y}_{\in}\sin\theta\sin\phi$) to solve the linearized kinetic equation~\eqref{lin-F}. 
The same values of initial hydrodynamic fields are used to solve linearized "hydrodynamic theories".

We close this subsection by noting that the rescaled solution $\e_{0}/\e_{\in}, \pi_{0}/\e_{\in}$ as well as the rescaled response function do not depend on the magnitude of $\e_{\in}$ and $\d \e_{\in},  g^{a}_{\in}$. 
Therefore, we shall not set their values in practice but simply present results in terms of rescaled quantities.

\begin{figure}[t]
    \centering
    \includegraphics[width=0.45\textwidth]{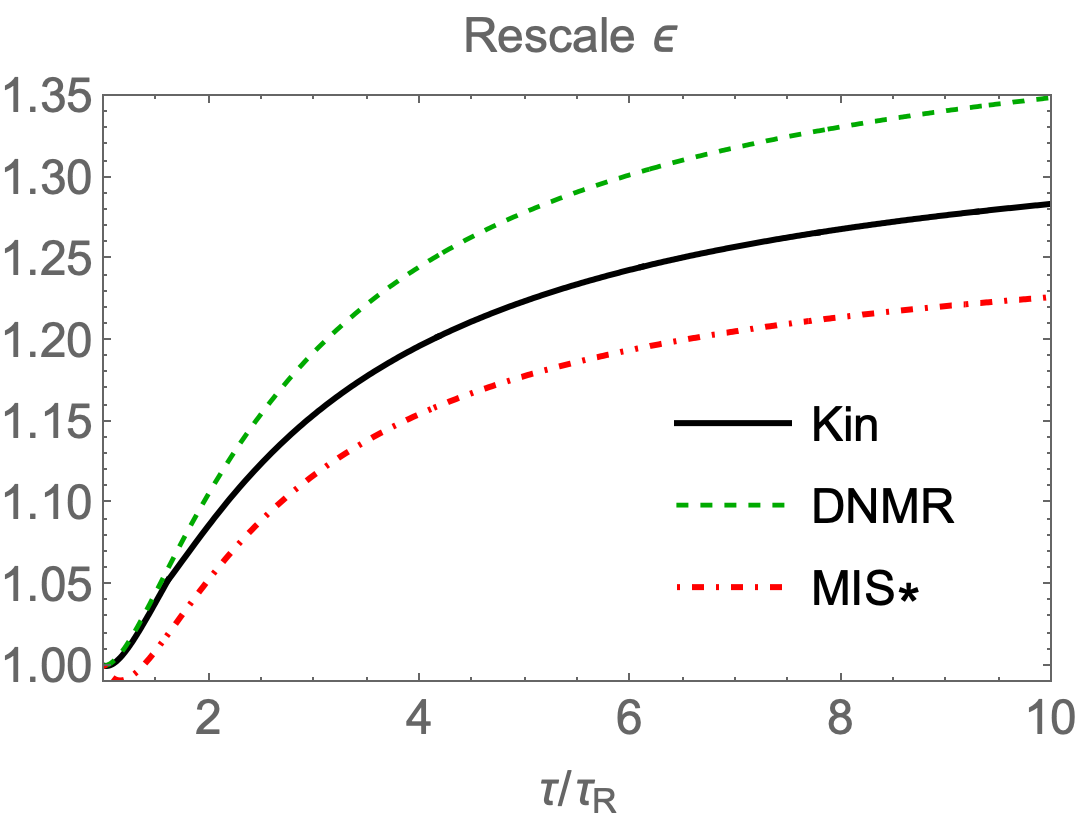}
    \includegraphics[width=0.45\textwidth]{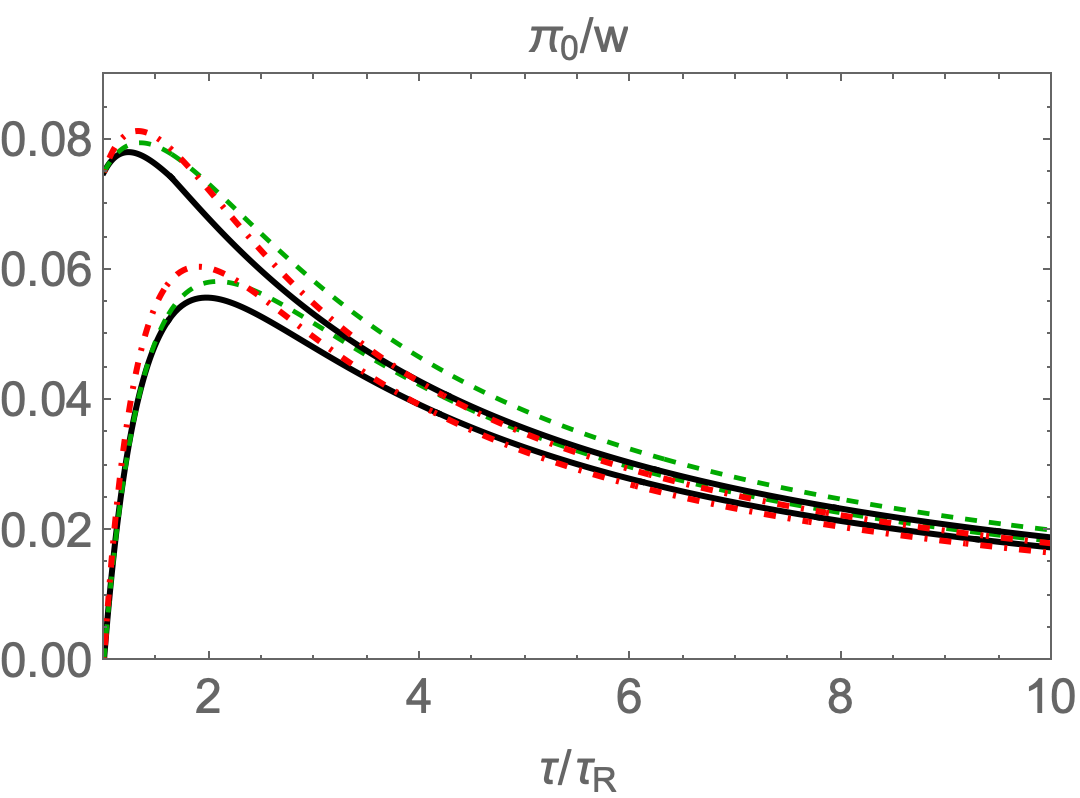}
    \caption{
    \label{fig:BG}
    Left: Background solution with $\pi_{\in}/w_{\in}=0$ for rescaled energy density $\e(\tau)/(\e_{\in}(\tau_{0}/\tau)^{4/3})$ from different "hydrodynamics theories" as compared to RTA solution. 
   Right: the evolution of $\pi_{0}/w$. The upper and lower curves correspond to the solutions with $\pi_{\in}/w_{\in}=0$ and $\pi_{\in}/w_{\in}=0.75$ respectively. 
    }
\end{figure}

\subsection{Background solution
\label{sec:B-background}
}

To determine the background evolution in different theories, 
we have solved the corresponding equations derived in Sec.~\ref{sec:kin-B} and Sec.~\ref{sec:hydro-B}.
In Fig.~\ref{fig:BG}
we plot the dimensionless ratio $\e_{0}(\tau)/(\e_{\in}(\tau_{0}/\tau)^{4/3}))$ (left)
and $\pi_{0}(\tau)/w_{0}(\tau)$ (right) as a function of the rescaled proper time $\tau/\tau_{R}(\e)$ for kinetic, DNMR and MIS* theories. 
For definiteness, MIS* results are obtained with the choice of MIS parameters $(\d,\g)=(0.2,0.8)$, as motivated by Fig.~\ref{fig:dispersion}. 
The solid and dashed curves correspond to $\pi_{\in}/w_{\in}=0$ and $\pi_{\in}/\e_{\in}=3/4$ respectively. 
We observe differences among different theories in the background evolution are within a few percent. 
This indicates that if we see any significant difference in response functions, it is mainly due to the discrepancy in their description of the response. 
Finally, Fig.~\ref{fig:BG} (right) where $\pi_{0}/w_{0}$ is plotted, we confirm that this ratio is indeed small, as we claimed earlier.
We also notice that background evolution is insensitive to the choice of $\pi_{in}/w_{\in}$ and we shall compute the response function for $\pi_{in}/w_{\in}$.

\subsection{Response in Fourier space}

\begin{figure}[t]
    \includegraphics[width=0.32\textwidth]{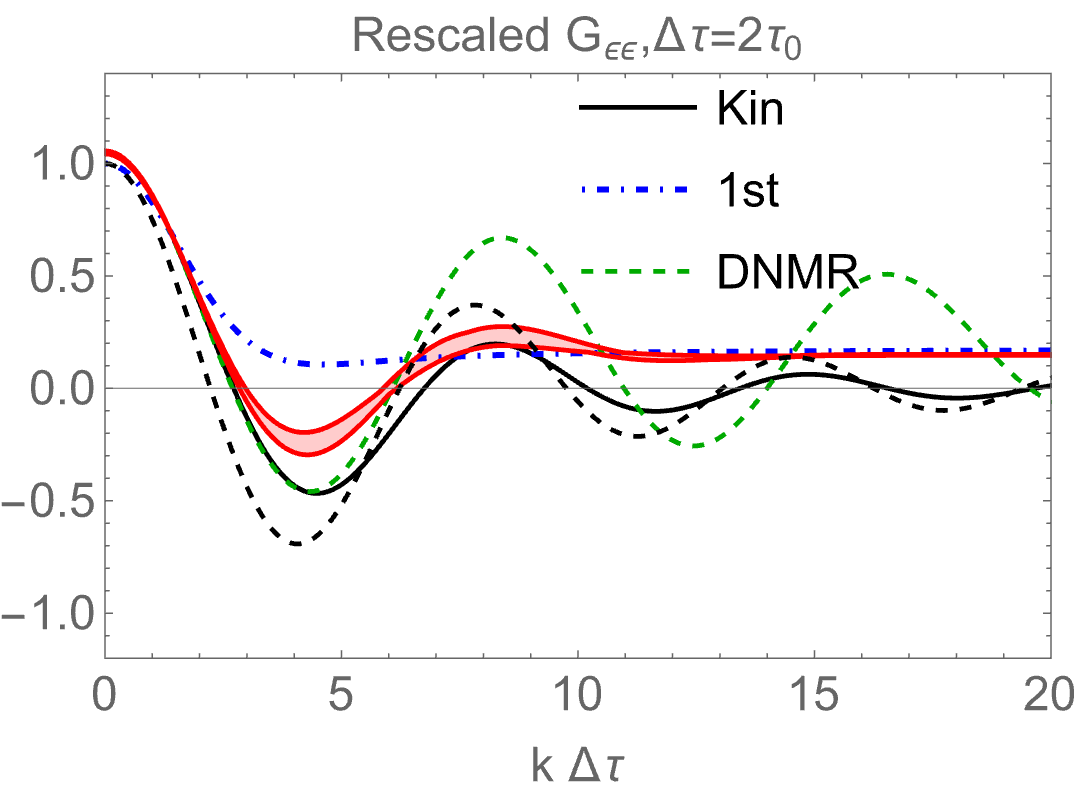}
    \includegraphics[width=0.32\textwidth]{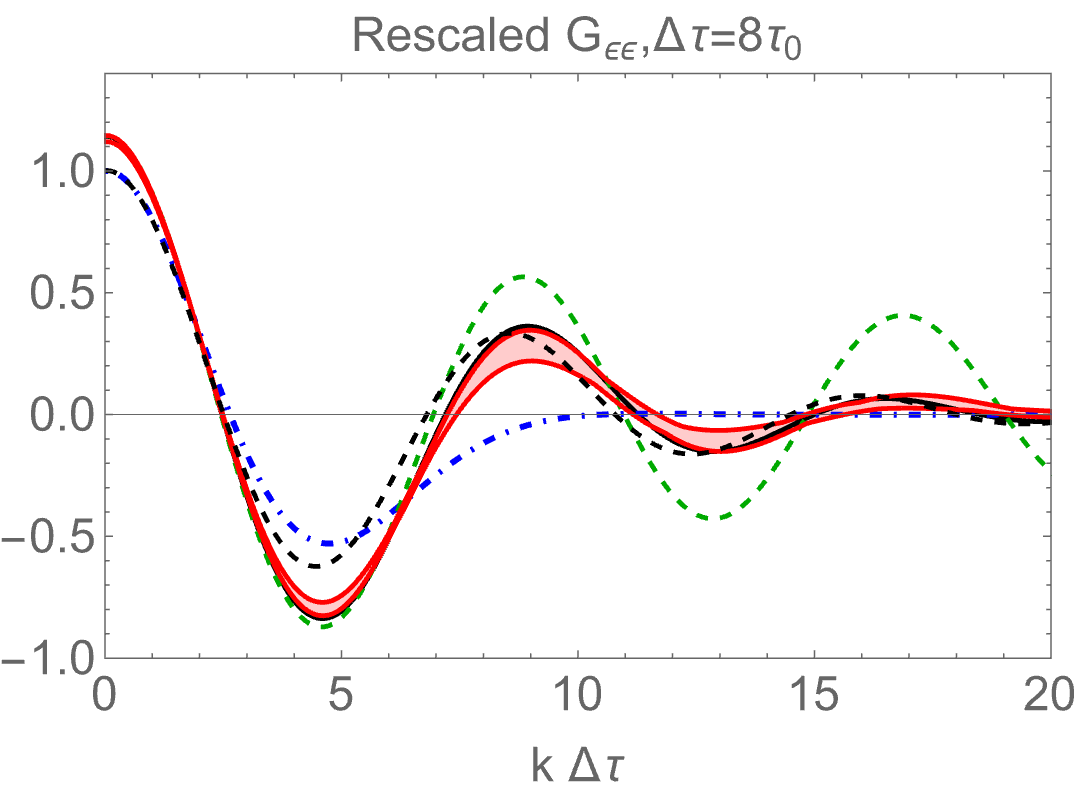}
    \includegraphics[width=0.32\textwidth]{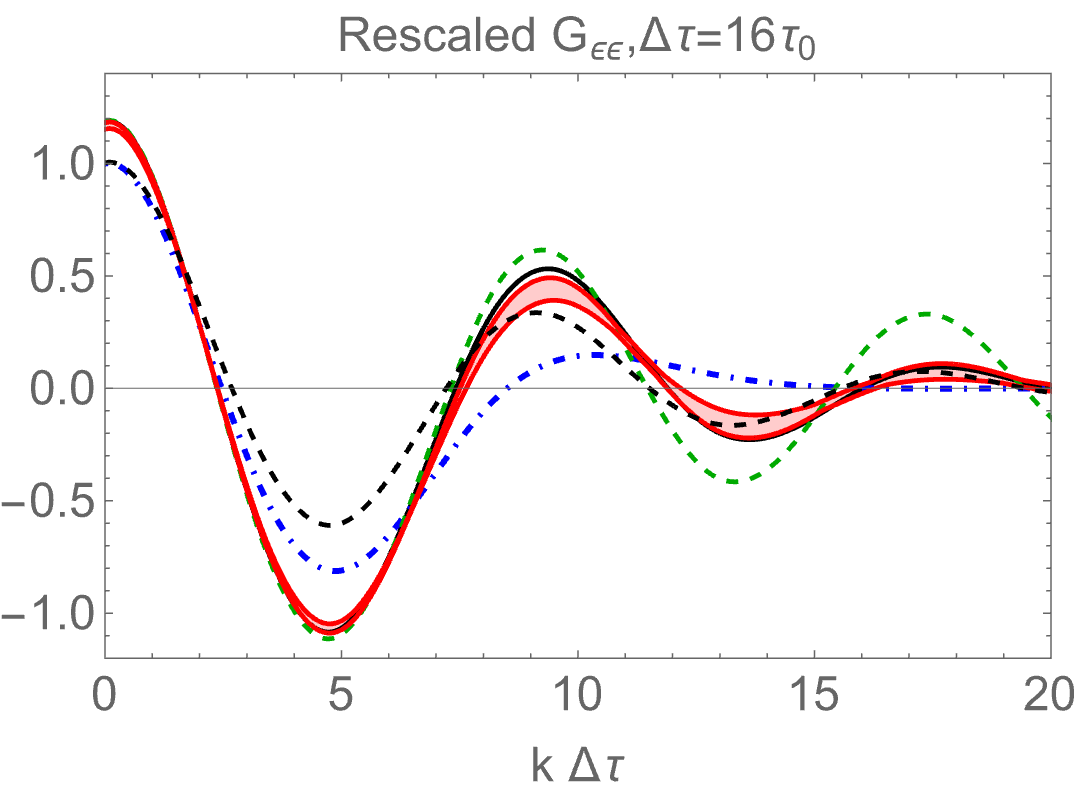}
    \includegraphics[width=0.32\textwidth]{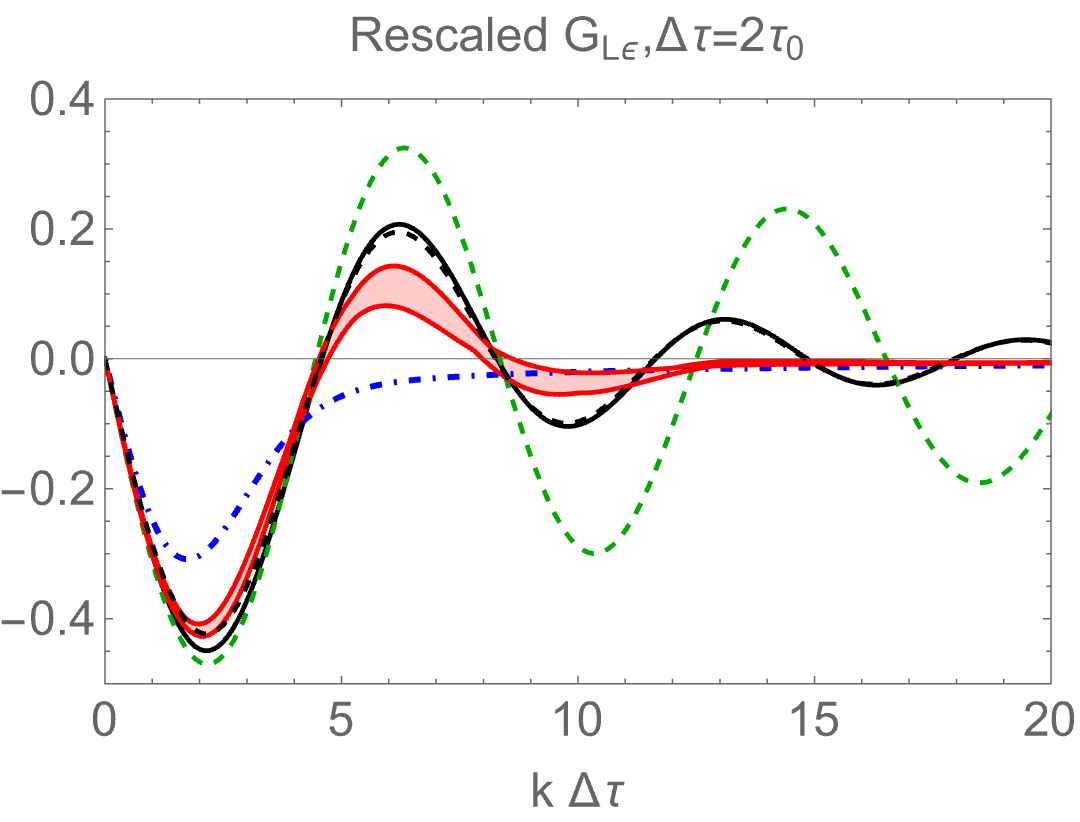}
    \includegraphics[width=0.32\textwidth]{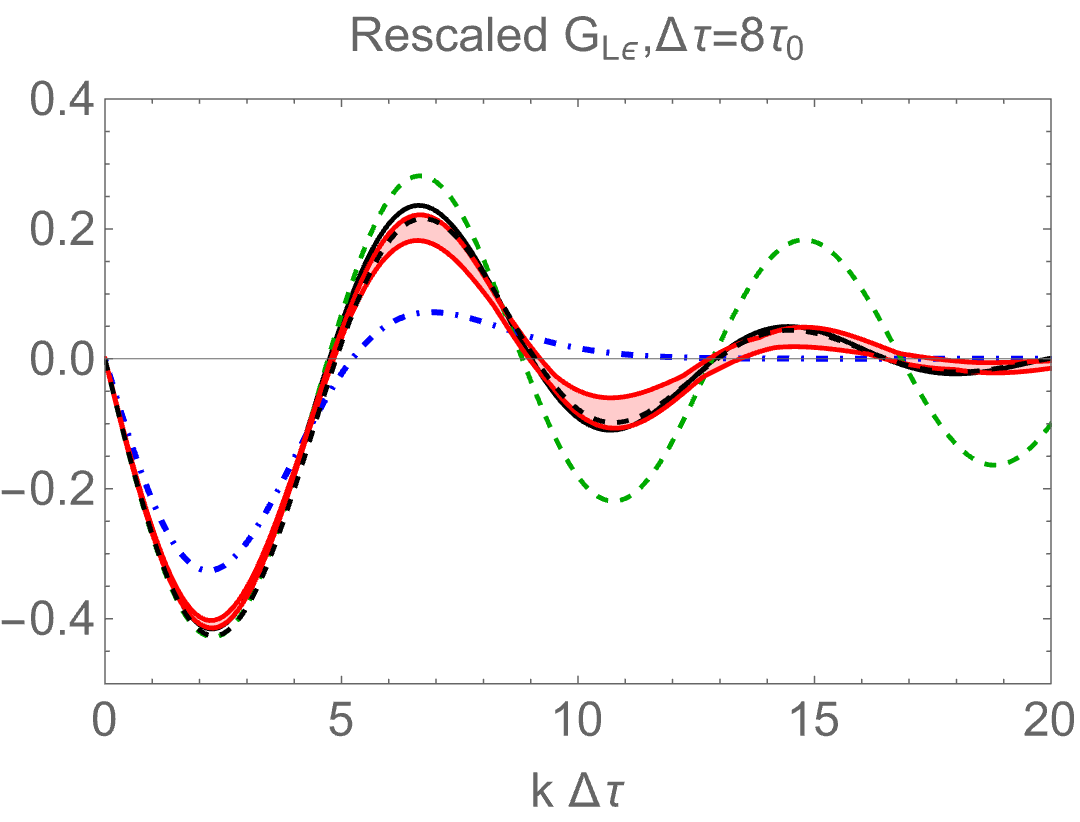}
    \includegraphics[width=0.32\textwidth]{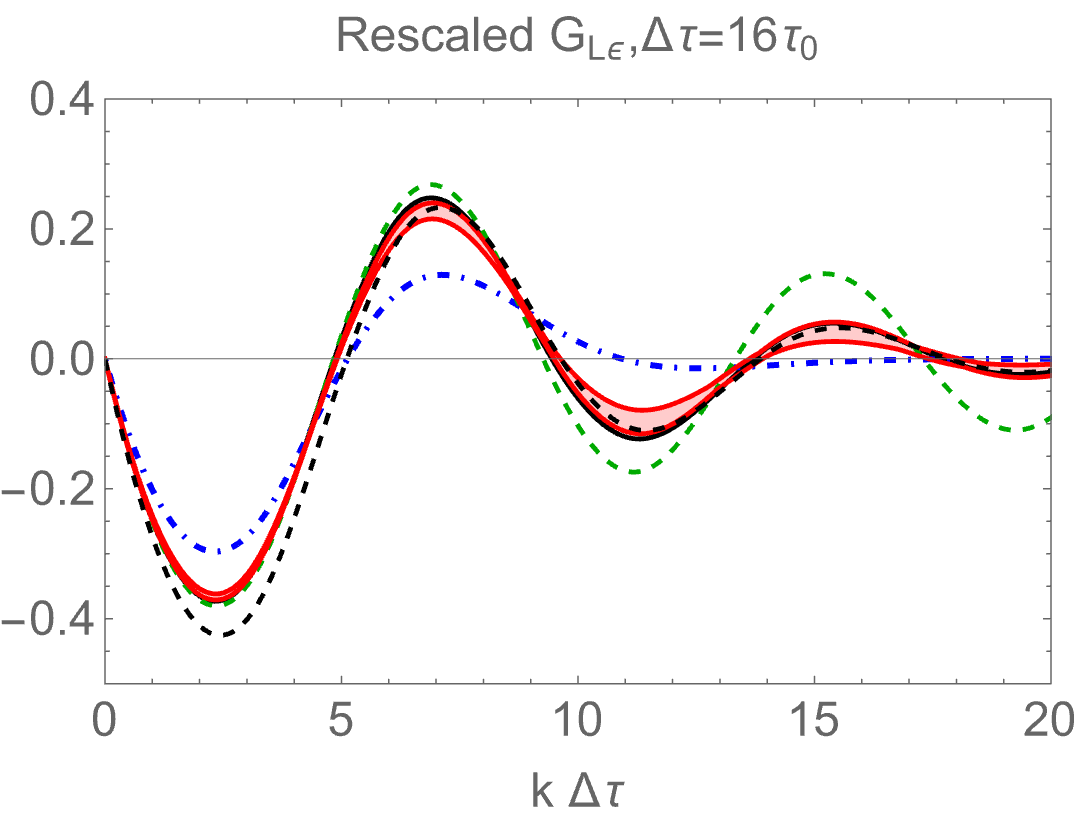}
    \includegraphics[width=0.32\textwidth]{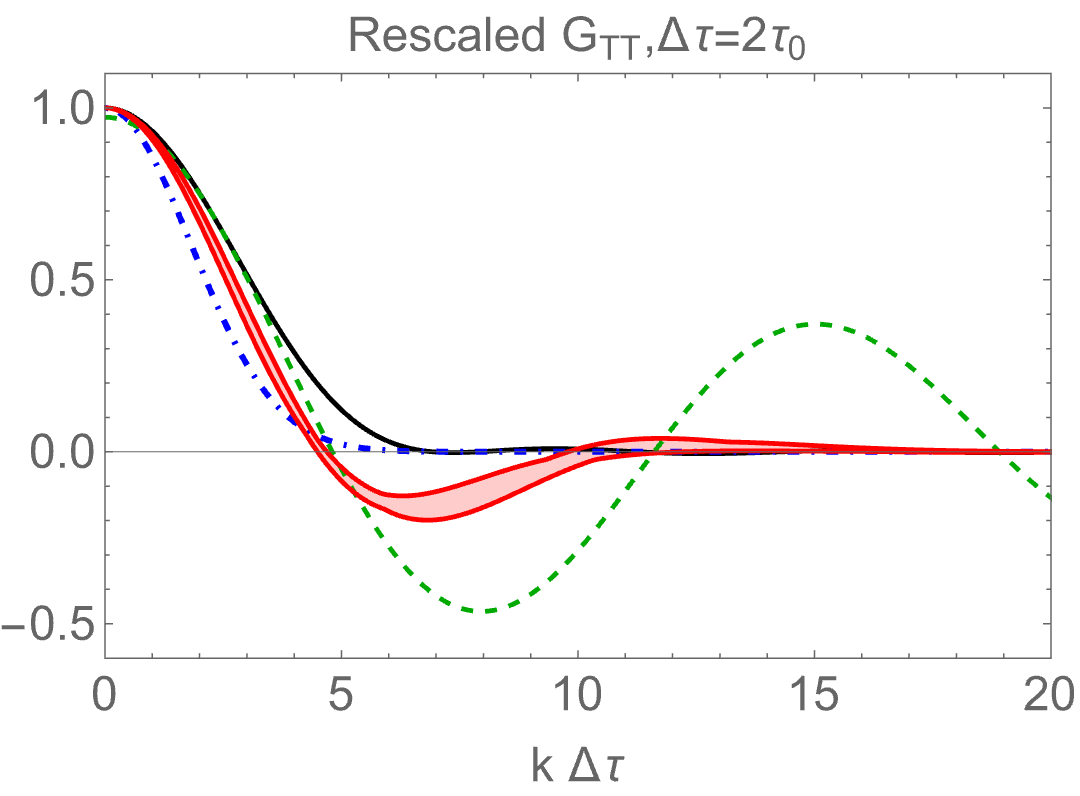}
    \includegraphics[width=0.32\textwidth]{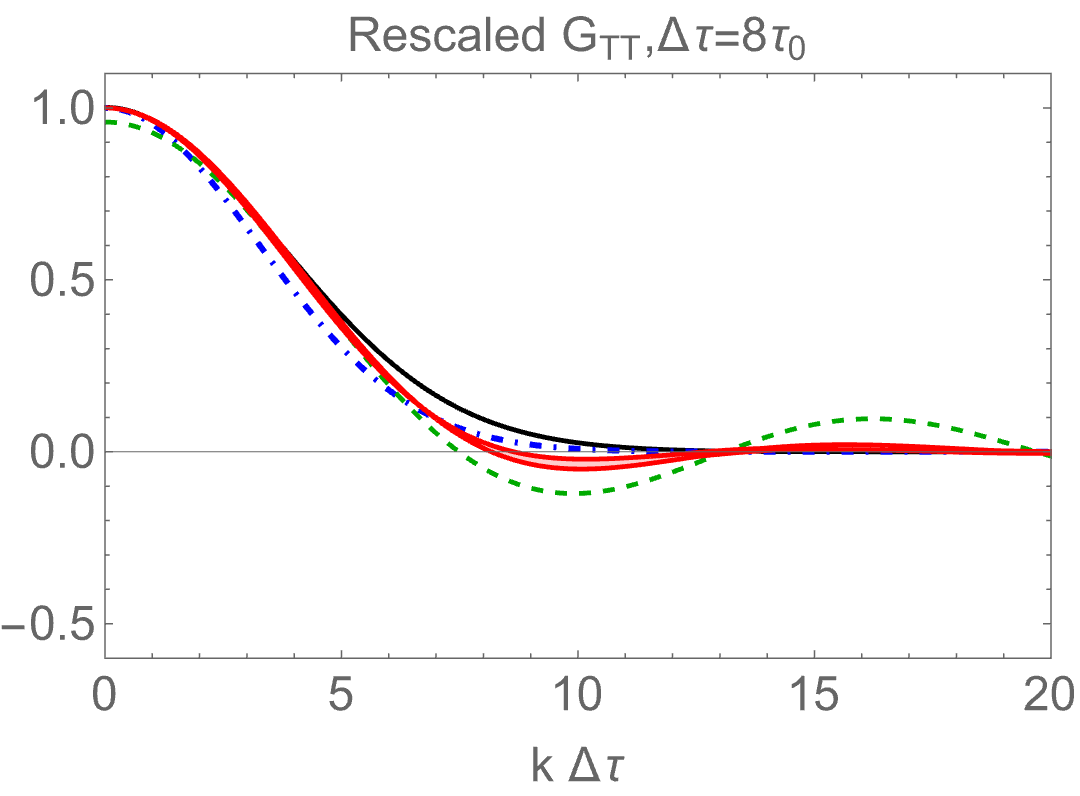}
    \includegraphics[width=0.32\textwidth]{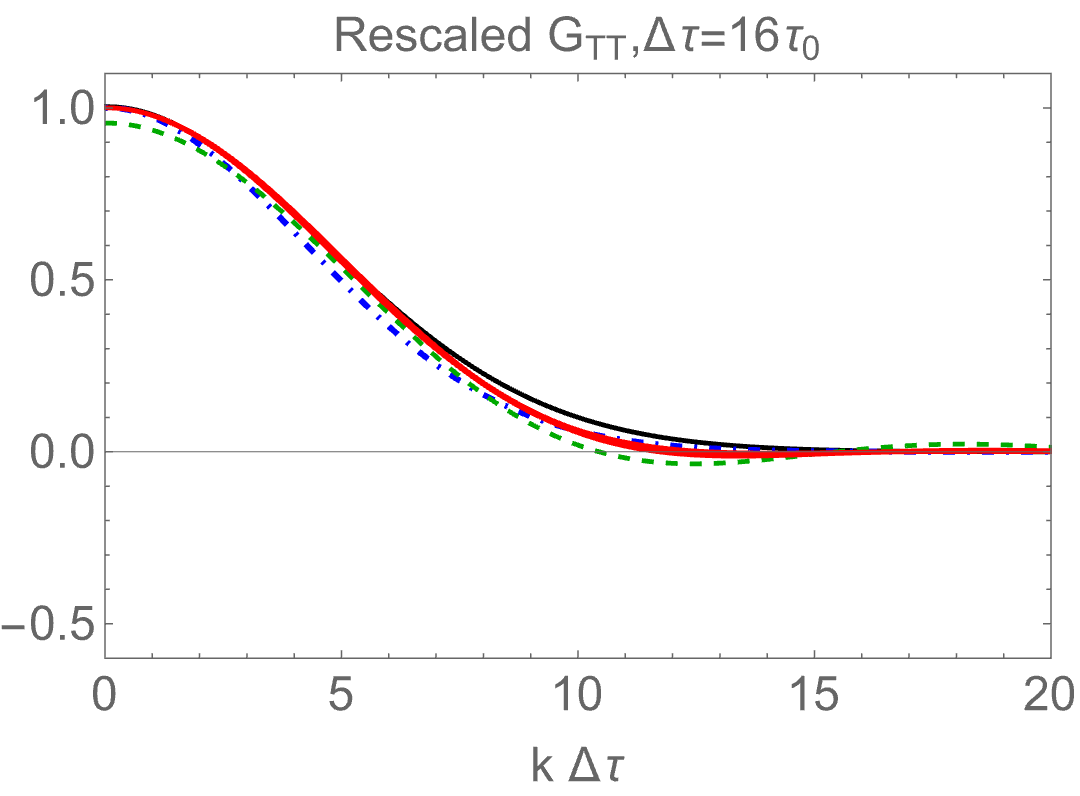}  
    \caption{
        \label{fig:G_vs_k}
The energy-momentum response function $G_{\e\e}$, $G_{L\e}$, $G_{TT}$, defined in eqs.~\eqref{RF-fourier} and \eqref{Gk-def-B} (c.f. their analogs in the static background in Fig.~\ref{fig:G_vs_k}), are plotted in the solid black curves from the first to the third row as a function of $k\Delta \tau$.
Those functions are rescaled by multiplying $(\tau/\tau')^{4/3},(\tau/\tau'), (\tau/\tau')$ respectively, see text.
The results of RTA kinetic theory, first-order hydrodynamics, and MIS theory are shown in the solid, dotted, and dashed curves, respectively. 
The MIS* results are plotted as red bands, computed from $0.2<\delta<0.3, 0.7<\g<0.8$. 
Three values of $\Delta\tau$ represent the response function at an early, intermediate, and late time (from left to right).  
To test relation~\eqref{G-relation-exp},
we plot $(\tau/\tau')G_{LL}$ and $(\tau/\tau')G_{\e L}/3$ from the kinetic theory in the black dashed curve in the first and second row, respectively.
They are to be compared with rescaled $G_{\e\e}$ and $G_{L\e}$. 
    }
\end{figure}

The functions defined in eq.~\eqref{G-list} describe the energy and momentum density distribution at $\tau$ caused by the energy-momentum perturbation at $\tau'$ for a Bjorken expanding plasma. 
Fig.~\ref{fig:G_vs_k} compares $G_{\e\e}, G_{\e L}, G_{TT}$ in spatial Fourier space in RTA kinetic theory (black curves) with those in "hydrodynamic theories."
When solving MIS*, we use the same range of $\delta, \gamma$ that gives a reasonable description of the RTA sound propagation in EHR. 
For better visualization, 
we rescale the response functions by multiplying $(\tau/\tau')^{a}$ so that their magnitudes are of the order of unity. 
The scaling exponent $a=4/3$ for $\sG_{\e\e}$ and $a=1$ 
for the others, guided by the evolution of the background energy and momentum density
for ideal hydrodynamics, which behave as $\tau^{-4/3}$ and $\tau^{-1}$ respectively.  
We present results with three representative values of $\Delta \tau=\tau-\tau'$, i.e. 
$\Delta \tau=2,8,16\tau_{0}$ in Fig.~\ref{fig:G_vs_k}.
Those plots reiterate that neither first-order nor MIS theory describes the response beyond the small gradient regime, while MIS* significantly improves the description to a larger gradient domain. 
Meanwhile, $\sG_{TT}(k)$, corresponding to the shear channel, "hydrodynamizes" at a smaller value of $\Delta \tau$ as compared with other response functions. 
This further emphasizes the significance of describing sound propagation for response at a non-hydrodynamic time scale.

When comparing the energy-energy response function $\sG_{\e\e}$ and the transverse momentum density response function $\sG_{TT}$ to their counterparts for the static background (i.e., $G^{L}$, $G^{T}$ in Fig.~\ref{fig:RTA-response}), an interesting observation can be made. 
The responses in both backgrounds are qualitatively similar and semi-quantitatively comparable. 
This can be explained by considering a fluid background with a characteristic size of in-homogeneity $l_{*}$ (This scale in a heavy-ion collision has been estimated in Ref.~\cite{Akamatsu:2016llw}  ). 
For gradients at $k >1/l_{*}$, including those living in the EHR, the response cannot resolve the long-wavelength structure of the background. 
Consequently, the EHR response should be insensitive to the difference in the background profile.

To better understand the preceding point, we simplify the first equation in eq.~\eqref{lin-B} under the limit $k\tau > 1$.
\begin{align}
\label{lin-B-EHR}
    \pd_{\tau}\delta \e + \partial_a\delta g^a = 0\, .
\end{align}
This equation is the same as the linearized energy conservation equation in the static background upon replacing the Bjorken time $\tau$ with $t$. 
Consequently, the response function for those two different backgrounds should be comparable.
Moreover, this also implies that in this regime, the number of independent \RF for the expanding background can be further reduced, as there are only two in energy-momentum response in static background. 
Indeed, we found in Fig.~\ref{fig:G_vs_k} that for Bjorken expansion, the following relation, which can be derived by taking $k\tau \gg 1$ limit, holds approximately
\begin{align}
\label{G-relation-exp}
   \left(\frac{\tau}{\tau_{0}}\right)^{4/3} G_{\e\e}\approx \frac{\tau}{\tau_{0}} G_{LL}
   \qquad
    G_{L\e}\approx \frac{1}{3} G_{\e L}\, .
\end{align}

\subsection{Response in real space}

\begin{figure}[t]
    \includegraphics[width=0.32\textwidth]{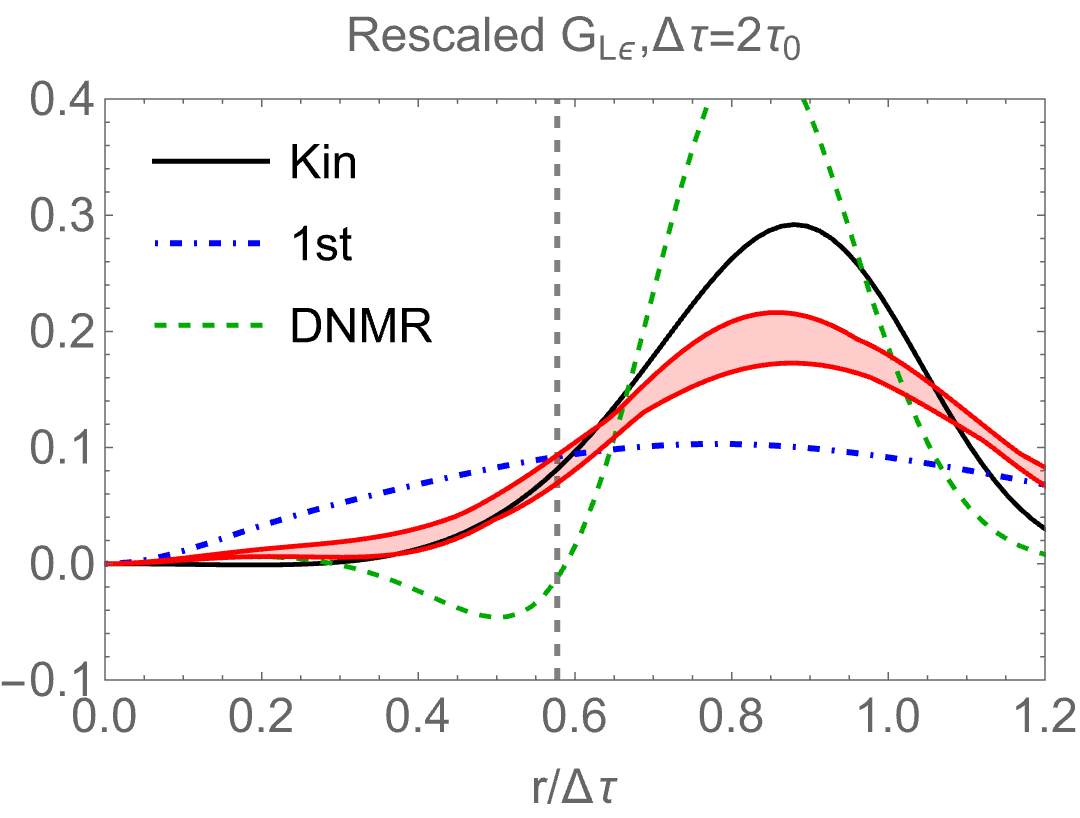}
    \includegraphics[width=0.32\textwidth]{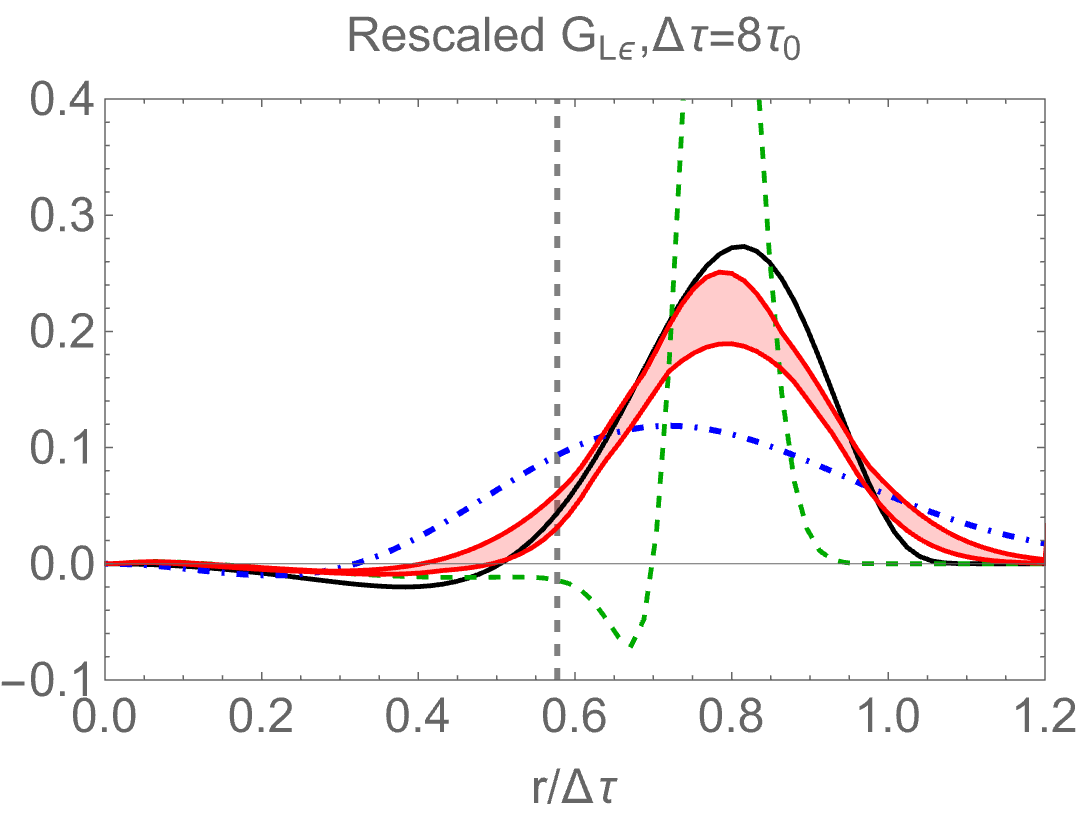}
    \includegraphics[width=0.32\textwidth]{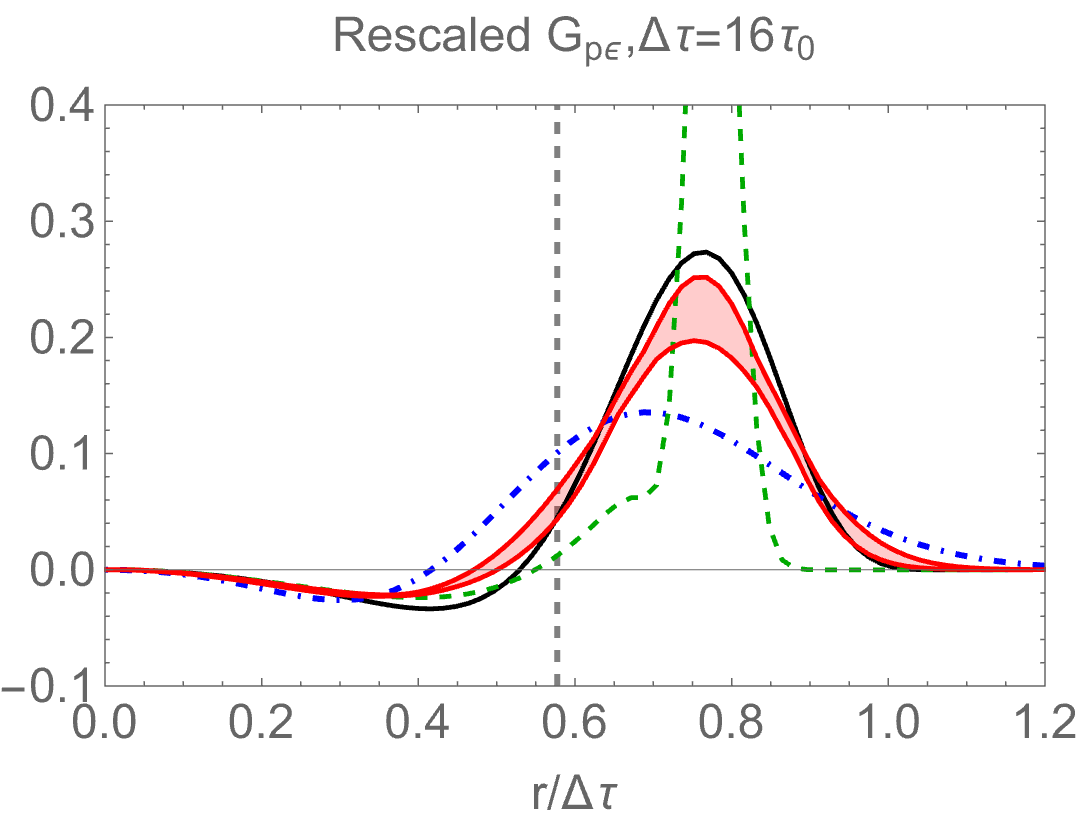}
    \includegraphics[width=0.32\textwidth]{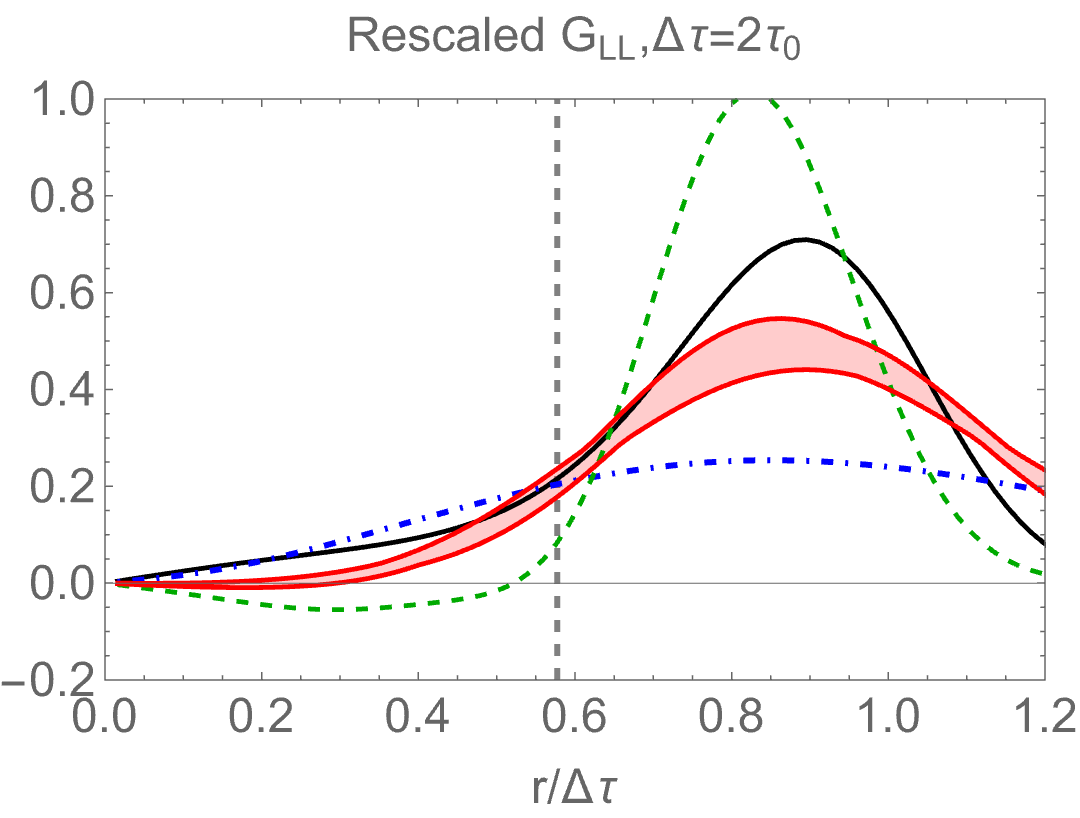}
    \includegraphics[width=0.32\textwidth]{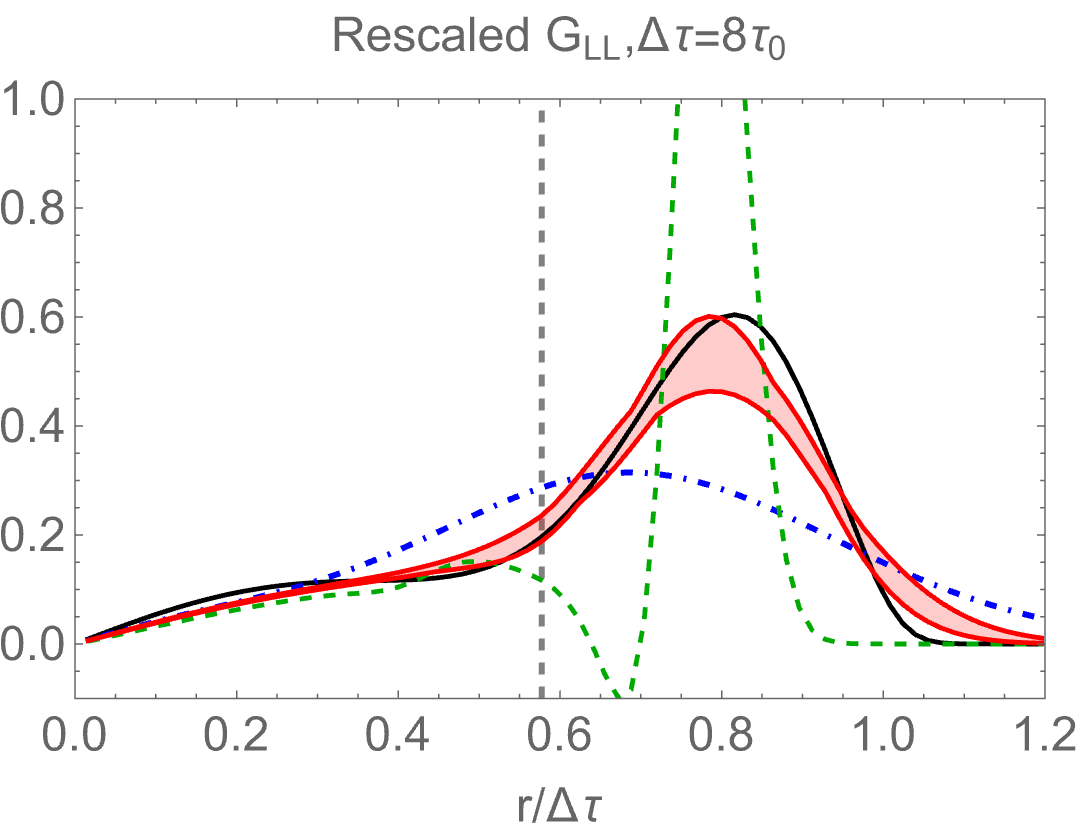}
    \includegraphics[width=0.32\textwidth]{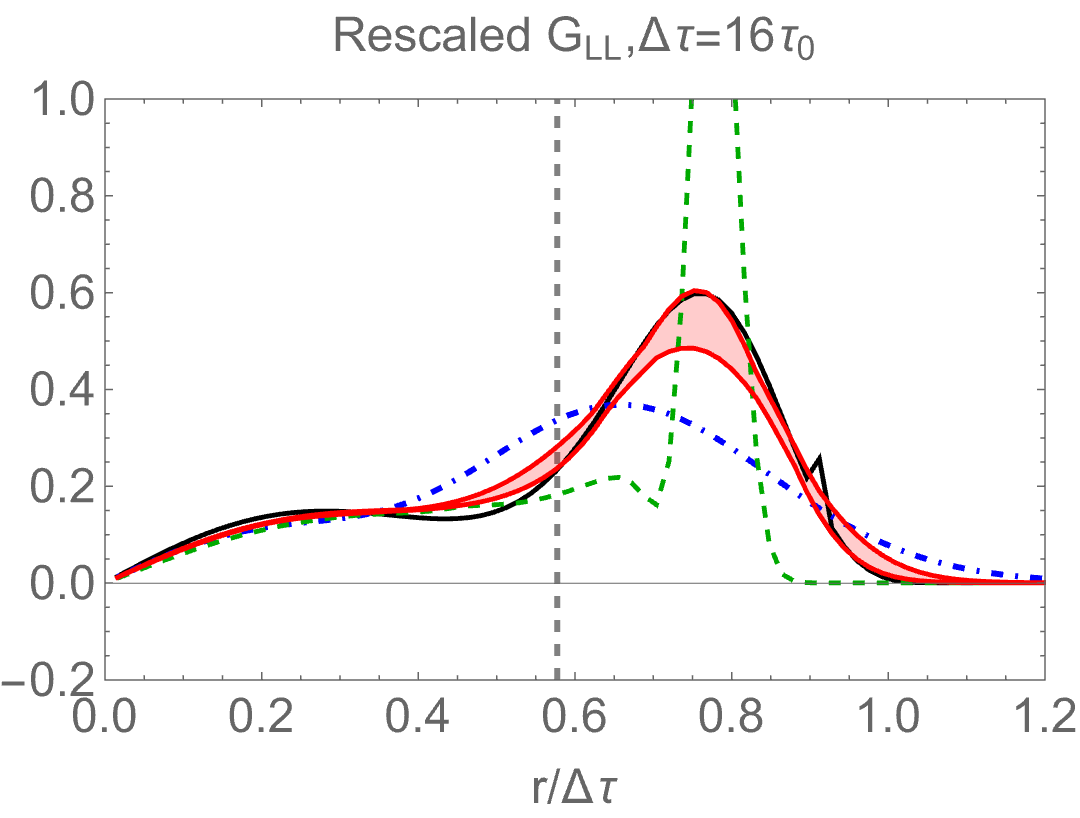}  
    \caption{
        \label{fig:G_vs_r}
The same as Fig.~\ref{fig:G_vs_k} but for the real-space response function $G_{L\e}, G_{LL}$ (defined in eq.~\eqref{G-list}) v.s $r/\Delta \tau$ for the Bjorken expanding plasma.
The vertical dashed black lines correspond to the value of conformal sound velocity $c_{s}=\sqrt{1/3}\approx 0.577$. 
    }
\end{figure}

We obtain the real space response function by taking the Fourier transforming according to eq.~\eqref{RF-fourier} where a smearing function $e^{-k^2/(2k_{\rm UV}^2)}$ in its integrand is included to eliminate contributions from high values of $k$. 
Introducing this UV cut-off is a common practice in previous studies~\cite{Kurkela:2018vqr}, 
and is necessary for several reasons. 
Firstly, we are interested in the response in EHR, which in general may not include arbitrarily large values of $k$
Secondly, the kinetic theory is invalid when the gradient is larger than the typical effective temperature, as the mixing between particle and anti-particle can not be ignored in such cases. 
Thirdly, in many physics situations, the source of energy-momentum perturbation should only have finite support in the $k$-space. 
In practice, we use $k_{\rm UV} = 3/\tau_{0}$ and have checked the results shown below are insensitive to the choice of $k_{\rm UV}$ as long as $k_{\rm UV}\gg\tau_{0}$.

In Fig.~\ref{fig:G_vs_r}, we have graphed the rescaled response function $\sG_{L\e}, \sG_{LL} $against $r/\Delta \tau$ for the same set of values of $\Delta \tau$ as we did in Fig.~\ref{fig:G_vs_k}(see Ref.~\cite{Ke:2022tqf} for $\sG_{\e\e}$). 
For causality to hold, \RF must disappear outside the causal circle $r/\Delta \tau \leq 1$. 
Due to the smearing function we introduce, the numerical real space \RF might be non-zero outside the causal circle, but it will be of a small magnitude.

The most important qualitative feature of the real space \RF associated with energy and momentum response (e.g., $\sG_{L\e}$ in the first row of Fig.~\ref{fig:G_vs_r}) is the presence of a peak at a specific value of $v_{\eff}=r/\Delta\tau$.
This value represents the effective velocity for the propagation of energy and momentum perturbation. 
The real space results are given by summing the contribution from different $k$ with an appropriate weight. 
Therefore $v_{\eff}$ also tells us the average phase velocity of the sound. 
As time passes, 
$v_{\eff}$ approaches the hydrodynamic limit $c_s = 1/\sqrt{3}$ at late times. However, even at $\Delta\tau=16\tau_R$, $v_{\eff}$ is still visually different from $c_s$.
Going back in time, $v_{\eff}$ is supersonic and becomes closer to $0.86$, the maximum value of the sound velocity in EHR (see Fig.~\ref{fig:dispersion}). 
This indicates that unless $\Delta \tau$ is very large, 
the contribution from EHR is significant, and high-frequency sound mode plays a crucial role in early times.

Regarding the peak's width, it measures the average sound attenuation rate, where a larger attenuation rate implies a broader peak. 
We anticipate from Fig.~\ref{fig:dispersion} and now confirm here that first-order theory (MIS/DNMR) will underestimate (overestimate) the peak width. 
In contrast, MIS* theory reasonably describes almost all the \RF functions under consideration with a suitable model parameter $\delta, \g$. 
This confirms that MIS* can be utilized to describe the response in the EHR. 

Finally, we note that $\sG_{LL}(r)$, defined in relation~\eqref{eq:response:Gk-Gr} and shown in the second row of Fig.~\ref{fig:G_vs_r},  receive contributions from both sound and shear modes, but it exhibits the sound peak in $r/\Delta \tau$ as its main feature. 
We have also checked that this peak is also in $\sG_{TT}$.
This further makes us believe that describing non-hydrodynamic sound propagation is more important than that for the shear mode.

\section{Non-hydrodynamic response from the Euclidean correlators
\label{sec:GE}
}

\begin{figure}
    \centering
    \includegraphics[width=0.32\textwidth]{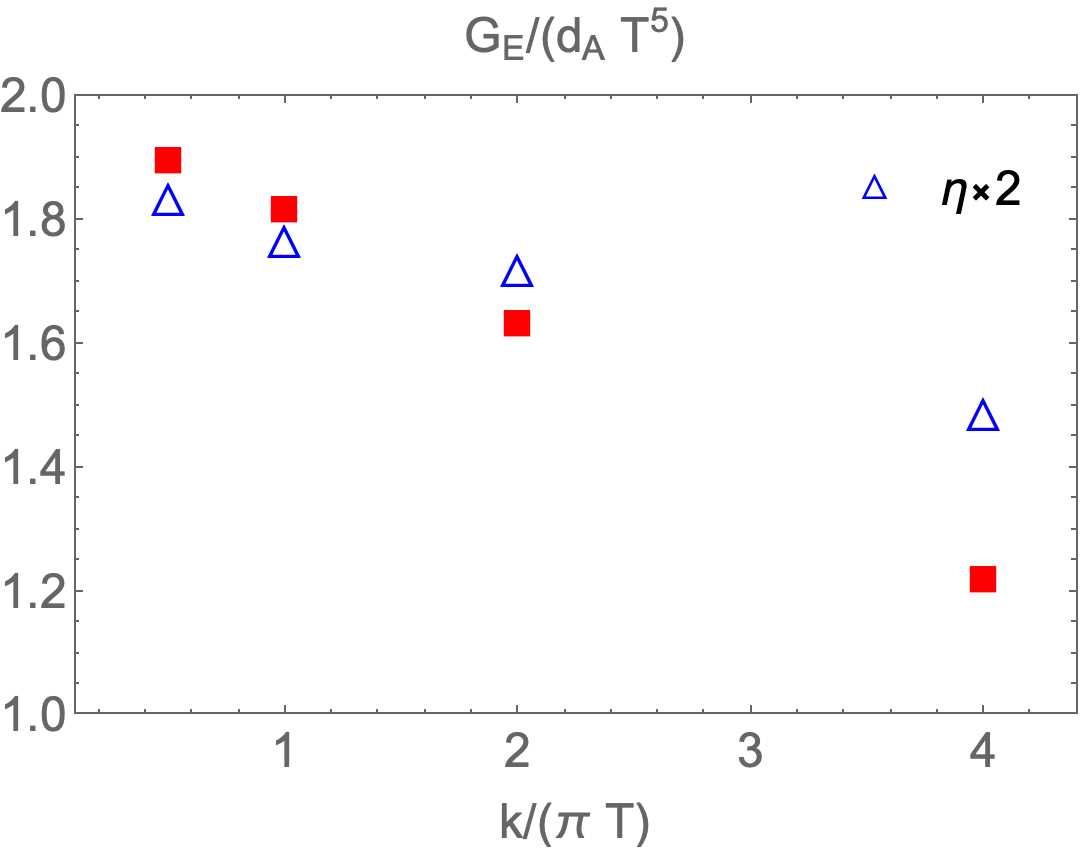}    
    \includegraphics[width=0.32\textwidth]{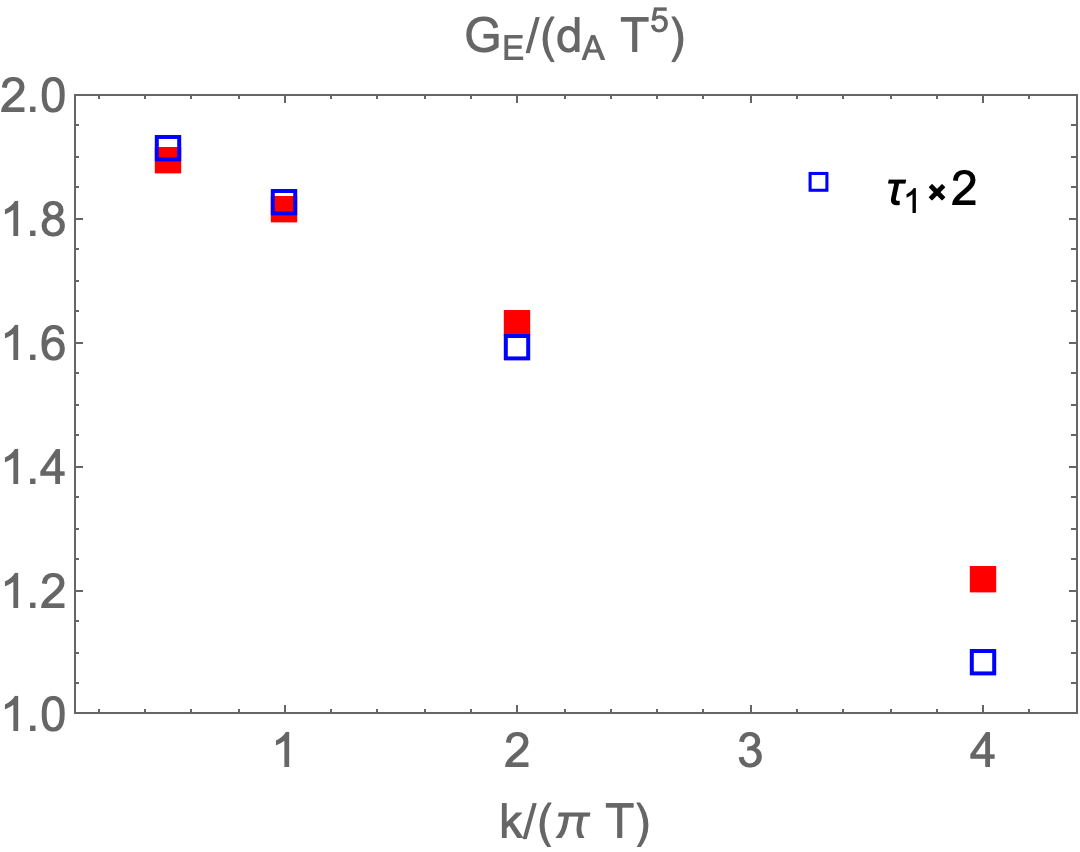}    
    \includegraphics[width=0.32\textwidth]{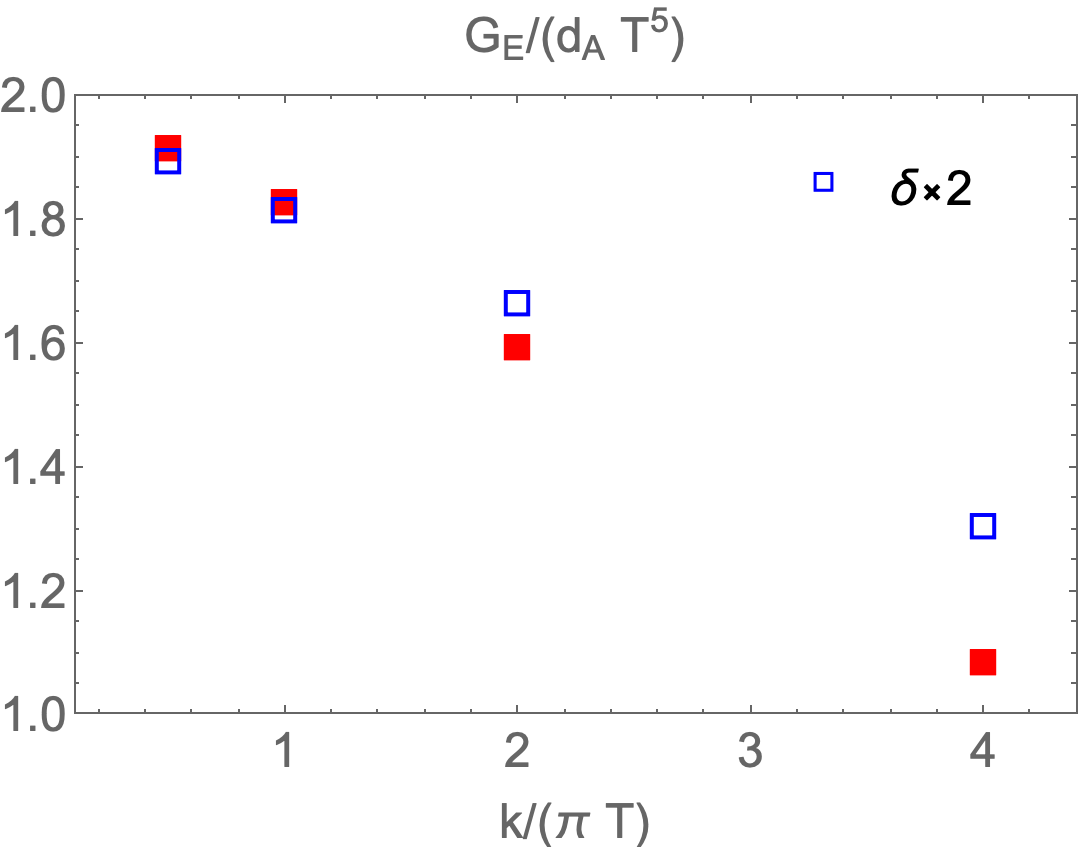}    
    \caption{
        \label{fig:GE}
    The rescaled Euclidean energy-energy retarded correlator $G_{E}$ computed using eqs.~\eqref{GE-rho} and~\eqref{rho-ansartz} at Euclidean time $\tau_{E}=1/(2T)$ for four different values of $k$. 
    The filled squares show results obtained with the benchmark values of parameters in the parameterization of spectral density in the extended hydrodynamic regime, $(\eta/s, \tau_{1},\delta)=(1/2\pi,4\eta/(sT),0.1)$. 
    They are compared with $G_{E}$ obtained from varying those parameters. See text. 
    }
\end{figure}

In this section, we discuss the implications of the EHR scenario for extracting the real-time properties of QGP from Euclidean correlators, which are directly accessible from first-principle lattice QCD calculations.

We recall the relation between Euclidean correlator $G_{E}$ and the spectral density $\rho$:
\begin{align}
\label{GE-rho}
    G_{E}(\tau_{E}, k)=
    \int^{\infty}_{0}d\o \rho(\o, k)\, \frac{\cosh\o\left(\tau_{E}-\frac{1}{2T}\right)}{\sinh\frac{\o}{2T}}\, . 
\end{align}
where $\tau_{E}$ is the Euclidean time. 
There have been extensive efforts in extracting the low-frequency behavior of spectral density from the lattice QCD calculation of $G_{E}$ to determine transport coefficients such as shear viscosity non-perturbatively (see Ref.~\cite{Meyer:2011gj} for a review).
Nevertheless, since the characteristic frequency of the kernel that is convoluted with the spectral density in eq.~\eqref{GE-rho} is of the order $\o_{T}=\pi T$, 
the resulting Euclidean correlators are notoriously insensitive to the features in hydrodynamic regimes where the typical gradient and frequency are expected to be smaller than $\o_{T}$.

Here, we argue that the characteristic gradient/frequency in the conjectured EHR of QGP should be of the order $\omega_{T}$ --- a frequency domain to which the Euclidean correlator may still be sensitive.
To support our claim, 
we consider the rescaled wave momentum $\tilde{k}= \frac{k}{(s/\eta)T c^{2}_{s}}$. 
While $\eta/s$ can be different even by order of magnitude for different microscopic theories, 
we hope that the dimensionless value of the characteristic $\tilde{k}$ in EHR, $\tk_{\EHR}$, is of the same order in those theories. 
In fact, 
the study of Ref.~\cite{Du:2023bwi} indeed indicates that the retarded correlator expressed in terms of $\tilde{k}$ is much less sensitive to the microscopic details. 
For RTA kinetic theory, we can use eq.~\eqref{H-para} and read from Fig.~\ref{fig:dispersion} that $\frac{9}{16}<\tilde{k}<\frac{9}{4}$ may be considered as the domain of EHR. 
In Ref.~\cite{Hong:2010at}, the spectral density of EMT has been calculated using the leading log Boltzmann equation for the QCD plasma. 
In sound channel, the authors observe that while hydrodynamics ceases to be a good description for $\tilde{k}>0.7$,
the peak associated with high-frequency sound (in the view of the present paper) in the spectral density persists up to $\tilde{k}\sim 5.6$ (see Fig.~2 of Ref.~\cite{Hong:2010at}). 
On the other hand, for the SYM correlator obtained from the AdS/CFT method, the EHR exists for $\tilde{k}>1$.  
Combining all the information above, 
it is reasonable to assume that for QGP, $\tk_{\EHR}=1\text{--}2 $, or
\begin{align}
    k_{\EHR}= (1\text{--}2)\, \left(\frac{\eta}{s}\right)^{-1}\, c^{2}_{s}\, T\, . 
\end{align}
Taking $\eta/s= 2\times (1/4\pi)$ as a crude estimate for QGP gives $k_{\EHR}\sim \pi T$.

We now investigate the sensitivity of the Euclidean correlator $G_{E}$ to the properties of EHR. 
For definiteness, we consider the energy-energy Euclidean correlator so that spectral density is given by $\rho= -{\rm Im}G^{00,00}_{R}/\pi$.   
Motivated by Refs.~\cite{Meyer:2008gt,Borsanyi:2018srz}, 
we assume a very simple EHR plus tree-level ansatz for the spectral function
\begin{align}
\label{rho-ansartz}
    \rho(\o,k)= \rho_{\EHR}(\o,k)\theta(k-\o)+\rho_{{\rm tree}}(\o,k)\theta(\o-k)\, , 
\end{align}
Here, $\rho_{\EHR}$ can be obtained by computing $G_{R}$ from the MIS* response function through the relation~\eqref{G-relation}. 
We find
\begin{align}
\label{rho-EHR}
    \rho_{\EHR}(\o,k)=\frac{-1}{\pi}{\rm }{\rm Im}\,\le[ \frac{w k^{2}}{\o^{2}-c^{2}_{s}k^2+i \nu(\o)\o k^{2}}\ri]\, , 
\end{align}
where we shall use the expression of $\nu(\o)$ in MIS*~\eqref{nu-KY2}, which we copy below for convenience
\begin{align}
\label{nu-KY20}
    \nu(\o)= \frac{4\nu_{0}}{3}\le(\frac{1-\delta}{1-i\o\tau_{1}}+\delta\ri)\, . 
\end{align}
In the estimation below, we always use $w=\frac{3}{4} w_{SB}=\frac{1}{15}d_{A}\pi^{2}T^{4}$ with $d_{A}=N^{2}_{c}-1$ and $w_{SB}$ denotes the enthalpy density of a gluon plasma in Stefan-Boltzmann limit. 
The tree-level spectral function of a gluon plasma is taken from Ref.~\cite{Meyer:2008gt}:
\begin{align}
    \rho_{{\rm tree}}(\o,k)\theta(\o-k)
    = \frac{d_{A}k^{4}}{4(4\pi)^{2}}\, 
    \int^{1}_{0}dz\, 
    \frac{(1-z^{2})^{2}\sinh(\frac{\o}{2T})}{\cosh(\frac{\o}{2T}-\cosh{\frac{k z}{2T}})}\, , 
\end{align}

We have calculated the rescaled Euclidean correlator $\tilde{G}_{E}\equiv G_{E}/(d_{A}T^{5})$ from eq.~\eqref{rho-ansartz} for $k=(1/2,1,2,4)\pi T$ at $\tau_{E}=\frac{1}{2 T}$. 
The results will depend on parameters entering eq.~\eqref{nu-KY20}, i.e. $\nu_{0}$ (or $\eta/s $) and MIS* parameters $\tau_{1}, \delta$. 
We first perform the computation with the benchmark values $(\eta/s, \tau_{1},\delta)=(1/2\pi,4 \eta/(s T), 0.1)$.
Note that for SYM, $T \tau_{\pi}=2.61 (\eta/s)$ while a next leading order QCD calculation show $T\tau_{\pi}\approx 6.5 (\eta/s)$~\cite{Ghiglieri:2018dgf}. 
So we pick up a somewhat arbitrary value in between for $\tau_{1}$ assuming $\tau_{1}$ and $\tau_{\pi}$ should be of the same order. 
The value of $\delta$ is inspired by the comparison between high-frequency sound dispersion in MIS* theory and that in RTA kinetic theory and SYM theory~\cite{Ke:2022tqf}. 
Then we do the same but increase the value of one of the parameters by a factor of two to explore the sensitivity of $G_{E}$ to those parameters.

The results shown in Fig.~\ref{fig:GE} are suggestive and encouraging. 
First, it confirms that at small $k$, the Euclidean correlator is insensitive to the value of the transport coefficient and MIS* parameters. 
For example, for $k=\frac{\pi T}{2}$ which might be considered as a gradient in the hydrodynamic regime, 
a factor of two difference in $\eta/s$  only results in a few percent difference in $G_{E}$. 
On the other hand, 
as $k$ increases, we observe sizable difference in $G_{E}$ with varying MIS* parameters $\tau_{1}, \delta$. 
Furthermore, since the energy-energy response in EHR is determined by MIS* parameters together with $\eta/s$, 
we also observe the sensitivity of $G_{E}$ to the value of shear viscosity at a non-hydrodynamic gradient. 
Those results suggest that by studying Euclidean correlators, we may be able to test the EHR scenario for QGP and could potentially extract both shear viscosity and parameters that characterize the medium's response in EHR.

\section{Summary and outlook
\label{sec:outlook}
}

We have studied the response of a QGP-like plasma to initial energy-momentum disturbance at a gradient outside the hydrodynamic regime.
By comparing RTA kinetic theories results with those obtained from "hydrodynamic theories," 
we demonstrate a naive extrapolation of the first-order and second-order (or MIS/DNMR theory) outside hydrodynamic regime that fails to describe the response. 
Nevertheless, high-frequency sound modes might dominate the response at a non-hydrodynamic gradient. 
Under this extended hydrodynamic regime scenario, the description of the response can still be simplified significantly. 
Notably, we observe the significant improvement achieved by MIS* that we propose.

We have also compared the energy-momentum response function at static and Bjorken-expanding. 
We learned that at a non-hydrodynamic gradient, the response as a function of proper time is not very sensitive to the bulk background profile.
This indicates that one may use the response functions obtained in the static background to estimate that in the expanding bulk profile, allowing for the simplification in the numerical studies. 
We also learned that although hydrodynamics would eventually describe those response functions, the "hydrodynamization time" of the response function would be numerically large. 
This, in turn, suggests that jet-medium observables might be employed to extract the medium's properties outside the hydrodynamic regime, and MIS* theory, or its extension, can be applied within this context.

We also hope the exploratory study presented in Sec.~\ref{sec:GE} would arouse interest in extracting the medium's behavior at a non-hydrodynamic gradient from the lattice Euclidean correlator. 
The latter should be sensitive to the response at the gradient of order $\pi T$. 
This is precisely the key missing gap in our understanding of QGP's behavior.

\acknowledgments

We thank Xiao-Jian Du, Aleksi Kurkela, Jia-Ning Li, Zong-lin Mo, Krishna Rajagopal, Soeren Schlichting and Xin-Nian Wang for useful discussions and comments.
Y.~Y. would like to acknowledge financial support by NSFC under grant No.12175282. 
W.~K. was supported by the LDRD Program at Los Alamos National Laboratory.
When preparing for the draft, we used Grammaly-Go AI to improve the presentation.

\bibliographystyle{JHEP.bst}
\bibliography{ref}

\end{document}